\documentclass[twocolumn,epjc3]{svjour3}
\usepackage{setup}
\usepackage{amsmath}
\usepackage{amssymb}
\usepackage{graphicx}
\usepackage{latexsym}
\usepackage[colorlinks,citecolor=blue,urlcolor=blue,linkcolor=blue]{hyperref}
\usepackage{cite}
\usepackage{ifthen}
\usepackage{subcaption}
\usepackage{siunitx}
\usepackage{enumitem}
\usepackage{booktabs}
\usepackage{appendix}
\newcommand\yZ{y^{\PZ}}
\newcommand\mgaNLO{{\sc\small MadGraph5\_aMC@NLO}\xspace}

\newcommand\PYTHIA{{\sc\small Pythia8}\xspace}
\newcommand\HERWIG{{\sc\small Herwig7}\xspace}

\journalname{Eur. Phys. J. C}

\begin{document}

\title{NNLO QCD predictions for Z-boson production in association with a charm jet within the LHCb fiducial region}

\author{%
  R.~Gauld\thanksref{MPI,rg}
  \and  
  A.~Gehrmann--De Ridder\thanksref{ETH,UZH,ag}
  \and
  E.~W.~N.~Glover\thanksref{DUR1,DUR2,ng}
  \and
  A.~Huss\thanksref{CERN,ah}
  \and
  A.~Rodriguez Garcia\thanksref{ETH,ar}
  \and
  G.~Stagnitto\thanksref{UZH,gs}
}

\thankstext{rg}{rgauld@mpp.mpg.de}
\thankstext{ag}{gehra@phys.ethz.ch}
\thankstext{ng}{e.w.n.glover@durham.ac.uk}
\thankstext{ah}{alexander.huss@cern.ch}
\thankstext{ar}{adrianro@phys.ethz.ch}
\thankstext{gs}{giovanni.stagnitto@physik.uzh.ch}

\institute{%
  Max Planck Institute for Physics, F\"ohringer Ring 6, 80805 M\"unchen, Germany
  \label{MPI}
  \and
  Institute for Theoretical Physics, ETH, CH-8093 Z\"urich, Switzerland
  \label{ETH}
  \and
  Department of Physics, University of Z\"urich, CH-8057 Z\"urich, Switzerland
  \label{UZH}
  \and
  Institute for Particle Physics Phenomenology, Durham University,  Durham DH1 3LE, UK
  \label{DUR1}
  \and
  Department of Physics, Durham University,  Durham DH1 3LE, UK
  \label{DUR2}
  \and
  Theoretical Physics Department, CERN, CH-1211 Geneva 23, Switzerland
  \label{CERN}
}

\date{MPP-2023-37, ZU-TH 12/23, IPPP/23/09, CERN-TH-2023-034}

\maketitle

\begin{abstract}
  We compute next-to-next-to-leading order (NNLO) QCD corrections to neutral
  vector boson production in association with a charm jet at the LHC. This
  process is studied in the forward kinematics at $\sqrt{s}=13$\,TeV, which may
  provide valuable constraints on the intrinsic charm component of the proton.
  A comparison is performed between fixed order and NLO predictions matched to a
  parton shower showing mutual compatibility within the respective
  uncertainties.  NNLO corrections typically lead to a reduction of theoretical
  uncertainties by a factor of two and the perturbative convergence is further
  improved through the introduction of a theory-inspired constraint on the
  transverse momentum of the vector boson plus jet system. A comparison between
  these predictions with data will require an alignment of a flavour-tagging
  procedure in theory and experiment that is infrared and collinear safe.
\end{abstract}  

\section{Introduction}
\label{introA}
The study of scattering processes that involve the direct production of
(heavy)-flavoured jets, i.e.\ those consistent with originating from charm
($\Pqc$) or bottom ($\Pqb$) quarks, in association with a leptonically decaying
vector boson is essential for collider physics phenomenology.
They form a major background for several Standard Model (SM) physics processes,
including the production of a Higgs boson in association with a gauge boson
where the Higgs boson decays into heavy-flavoured
jets~\cite{ATLAS:2018kot,CMS:2018nsn,ATLAS:2019yhn,ATLAS:2020fcp,ATLAS:2020jwz},
as well as signals expected in models of physics beyond the SM
(BSM)~\cite{CMS:2019zmd,ATLAS:2021yij}.
Furthermore, they can provide unique information on the distribution of
flavoured partons inside the
proton~\cite{ATLAS:2014jkm,LHCb:2021stx,CMS:2022bjk}.
Focussing on the process of $\PZ$ plus flavoured jet at the Large Hadron
Collider (LHC), several measurements have been performed by the ATLAS, CMS and
the LHCb collaborations at 7 and 8\,TeV proton-proton collision
energies~\cite{ATLAS:2014rjv,CMS:2014jqj,LHCb:2014ydc,CMS:2017snu,CMS:2016gmz,ATLAS:2020juj}.
Recent studies at 13\,TeV by the CMS collaboration~\cite{CMS:2020hmf} presented
measurements of observables related to the production of $\Pqc$ and/or
$\Pqb$-quark jets in a sample containing a $\PZ$-boson produced in association
with at least one jet.

The production of a leptonically decaying $\PZ$-boson in association with a
charm jet, and particularly at forward kinematics~\cite{Boettcher:2015sqn} which
is the focus of this work, could yield a unique probe of the charm content of
the proton~\cite{Boettcher:2015sqn,Bailas:2015jlc,Lipatov:2016feu}, provided
that precise predictions and measurements of flavour-sensitive $\PZ+\cjet$
observables are available and can be compared at a similar level of accuracy.
The LHCb collaboration has recently analysed events containing a $\PZ$-boson and
a charm jet in the forward region of phase space in proton-proton
collisions~\cite{LHCb:2021stx}.  These measurements simultaneously provide
direct access to the small- and large-$x$ regions of the $\Pqc$-quark parton
distribution function (PDF) that is not well explored by other experiments.
Specifically, LHCb has presented results~\cite{LHCb:2021stx} for the ratio of
production cross sections $R^{c}_{j} = \sigma(\PZ+\cjet)/\sigma(\PZ+\jet)$. This
ratio is measured differentially as a function of the rapidity of the
$\PZ$-boson $\yZ$ in the range $ 2.0< \yZ <4.5$.
The experimental result for the ratio $R^{c}_{j}$ has been compared with several
SM predictions obtained at NLO QCD accuracy interfaced with a parton shower
(NLO+PS), each using different input PDF sets.
It is demonstrated that the most forward $\yZ$ region is particularly sensitive
to the theoretical modelling of the charm quark PDF in these sets, with the best
agreement between theory and data obtained by choosing a PDF set with a
valence-like intrinsic (non-perturbative) charm quark component.
The presence or absence of this component is a long standing theoretical
issue~\cite{Brodsky:1980pb,Brodsky:2015fna}.
Recently, the NNPDF collaboration has claimed evidence for an intrinsic charm
quark component in the proton~\cite{Ball:2022qks}, with a local significance at
the 2.5$\sigma$ level for momentum fractions in the region $0.3 \lesssim x
\lesssim 0.6$.
By including the LHCb data for $R^{c}_{j}$ in a reweighting of their fit, and
adopting a theory prediction based on NLO+PS, the local significance further
increases to about 3.0$\sigma$.
Other PDF fitting collaborations have independently investigated the possible
presence of intrinsic charm in the proton~\cite{Hou:2017khm}: for instance, a
recent analysis by the CTEQ-TEA collaboration~\cite{Guzzi:2022rca} concludes
that finding evidence for nonperturbative charm continues to be elusive, by
highlighting challenging aspects that must be confronted in extracting
nonperturbative charm in PDF fits.

State-of-the-art predictions for this kind of processes featuring the associated
production of a vector boson with one or more flavoured jets has reached
next-to-next-to-leading order (NNLO) accuracy in QCD calculations with massless
quarks~\cite{Gauld:2020deh,Czakon:2020coa,Hartanto:2022qhh}.
Given the importance of the LHCb data for PDF extractions, it is highly
desirable to have fixed-order predictions for the $\PZ+\cjet$ process in the
forward region, in order to incorporate data for $R^{c}_{j}$ or other
flavour-sensitive observables in global PDF analyses based on collinear
factorisation.
In this paper we focus on a description of $\PZ+\cjet$ production within the
LHCb fiducial region following the approach of~\cite{Gauld:2022lem} to define
flavoured jets, and provide a detailed comparison of (new) fixed-order
predictions up to \NNLO QCD accuracy as well as those based on \NLOPS for a
variety of differential distributions.

Despite our focus on the forward kinematics relevant to the LHCb detector, in
this work we refrain from performing a comparison to the available LHCb
data~\cite{LHCb:2021stx}.
This is due to a significant contamination of the observable $R^{c}_{j}$
measured by LHCb from Multiple Particle Interactions (MPI), which should be
removed/subtracted before considering this data in a (single parton scattering)
collinear PDF fit.
Moreover, the experimental definition of jet flavour in~\cite{LHCb:2021stx} is
not infrared and collinear (IRC) safe, rendering a massless fixed-order
calculation ill defined for the experimental set-up.
With the goal in mind to provide constraining information on the potential
presence of an intrinsic charm quark component within the proton, it is critical
that the definition of the presented data is IRC safe such that a massless
calculation (where collinear factorisation for the charm-quark PDF has been
performed) can be applied.
We elaborate more on these issues in Appendix~\ref{sec:irc} and~\ref{sec:mpi}.

The structure of the paper is as follows. In Section~\ref{sec:calc}, we present
the main ingredients entering the computation of observables associated with
$\PZ+\cjet$ production both at pure fixed-order perturbative QCD as well as
using the NLO+PS framework.
We further comment on the proposal of~\cite{Gauld:2022lem}, the {\em flavour
  dressing} algorithm, which allows flavour to be assigned to anti-$k_{\rT}$
jets in an IRC safe way.
In Section~\ref{sec:numerics}, after having described the numerical set-up and
defined the scale variation prescriptions, we present for the first time
fixed-order predictions up to next-to-next-to-leading order (NNLO) in QCD for
several observables related to the $\PZ+\cjet$ process computed for the LHCb
experimental fiducial region at 13\,TeV.  We compare these fixed-order
predictions with NLO predictions matched to a parton shower at the parton level
using the flavour dressing procedure to define flavoured jets. We further
investigate the impact of an additional constraint on the transverse momentum of
the vector boson plus jet system in the computations.
We shall find that the inclusion of this theoretically motivated cut, brings
NNLO and NLO+PS predictions closer and improves the perturbative convergence of
the fixed-order results for a large fraction of the flavour-sensitive
observables and for most of the kinematical range studied.
We also present predictions for the ratio $R^{c}_{j}$.
In Section~\ref{sec:concl}, we summarise our findings and discuss the prospects
of a direct comparison between theory and data in the future.
The Appendices contain a discussion of the IRC safety of the current
experimental definition (Appendix~\ref{sec:irc}) and the role of MPI in the
current experimental set-up (Appendix~\ref{sec:mpi}).

\section{Details of the calculation} 
\label{sec:calc}

In this paper, one of the main goals is to present fixed-order predictions
including QCD corrections up to ${\cal O}(\alpha_{s}^{3})$ for observables
related to the process $\Pp \Pap \to \Pl \Pal +\cjet + X$, yielding a final
state that contains a pair of charged leptons and at least one identified charm
jet. 

The computation of higher order corrections to observables with identified
flavour poses several challenges as compared to flavour-blind cross sections: it
requires a complete flavour tracking of the particles in all subprocesses (which
are inputs to the flavour-dependent jet reconstruction/tagging), and
additionally to those appearing in all subtraction terms (for a calculation
based on subtraction).
A flavour tracking procedure at parton level has been pioneered for the
computation of \bjet flavoured observables in~\cite{Gauld:2019yng}
and~\cite{Gauld:2020deh} and implemented within the \nnlojet parton-level
generator that is used here. Within this framework, which employs the antenna
subtraction method to capture the infrared behaviour of matrix-elements yielding
fully differential cross section predictions at NNLO level, the flavour tracking
procedure has the crucial property that it can be applied to any flavour-blind
computation already present in the \nnlojet code. For example, the computation
of $\PZ+\bjet$ observables in~\cite{Gauld:2020deh} relied on the use of the
existing flavour blind $\PZ+\jet$ computation presented
in~\cite{Gehrmann-DeRidder:2015wbt} and used the flavour-$k_{\rT}$ algorithm to
select \bjets.
To compute observables related to the $\PZ+\cjet$ production process in this
work, we adopt a similar strategy by using the $\PZ+\jet$ computation including
up to ${\cal O}(\alphas^3)$ corrections, and then apply the flavour tracking
procedure as was done for the $\PZ+\bjet$ process.

As was the case in~\cite{Gauld:2020deh}, the prediction of scattering processes
with heavy-flavour jets can be further improved by exactly including the
contribution from a massive heavy-flavour quark at fixed order---i.e.\ the
resultant prediction is made in a general mass variable flavour number scheme.
We follow a similar procedure here and include mass corrections up to ${\cal
  O}(\alpha_{s}^{2} m_{\Pqc})$ in the fixed-order distributions which are
labelled as NLO and NNLO.

A further complication which is encountered in a calculation of the type
presented here---the calculation of QCD corrections to a scattering process with
flavoured jets based on massless quarks---is that an IRC safe definition of jet
flavour must be used.
A first solution to this problem was introduced in~\cite{Banfi:2006hf}, with the
formulation of the IRC safe flavour-$k_{\rT}$ algorithm. This algorithm features
a $k_{\rT}$-like clustering sequence, and introduces a specific
flavour-dependent clustering sequence to achieve IRC flavour safety.
However, since the algorithm requires the knowledge of the flavour of all the
particles in the event at each step of the clustering, it is challenging to
realise experimentally, and so far has not been implemented in experimental
analyses.
Therefore, the kinematics and flavour of the jets obtained with the
flavour-$k_{\rT}$ algorithm are not compatible with those obtained in
experiment.
Various alternatives to the use of flavour $k_T$ in theoretical predictions have
recently been proposed~\cite{Caletti:2022glq,Caletti:2022hnc,Czakon:2022wam}.

In the present analysis, we will adopt the flavour dressing algorithm
of~\cite{Gauld:2022lem}.
This approach is particularly suitable as it enables to assign flavour quantum
numbers to a set of flavour blind jets, obtained with any jet clustering
algorithm.
This allows us to apply it to jets reconstructed with the anti-$k_{T}$
algorithm~\cite{Cacciari:2008gp}, i.e.\ the same algorithm which is used in
experiment to define the kinematics of the jets (although the flavour assignment
procedure is different, as detailed in Appendix~\ref{sec:irc}).
Here, we stress the fact that in the flavour dressing algorithm the flavour
assignment is entirely factorised from the initial jet reconstruction, hence the
kinematics of the jets is not affected.
Such a key property is relevant in the present context, since it ensures that
for a ratio observable such as $\sigma(\PZ+\cjet)/\sigma(\PZ+\jet)$ both the
numerator and the denominator feature the same sample of anti-$k_{\rT}$ jets.

A comparison of the fixed-order predictions as described above will also be made
to several NLO predictions matched with a parton shower.  Those NLO+PS
predictions are obtained either with the \mgaNLO
(v.\ 2.7.3)~\cite{Alwall:2014hca} framework interfaced to \PYTHIA
(v.\ 8.243)~\cite{Sjostrand:2014zea} (default $p_{\rT}$-ordered parton shower)
or \HERWIG (v.\ 7.2.2)~\cite{Bahr:2008pv,Bellm:2015jjp,Bellm:2019zci} (default
angular-ordered parton shower), and using the same flavour dressing procedure to
define flavoured jets as for the fixed-order predictions.
To allow for a more direct comparison, those predictions are obtained at the
parton level where neither the impact of MPI or hadronisation effects are
included.
A discussion on the important role of MPI for the $\PZ+\cjet$ process in the
forward region is provided in Appendix~\ref{sec:mpi}.  Hadronisation effects, on
the other hand, were found not to impact the considered observables in any
significant manner.

\section{Numerical results} 
\label{sec:numerics}

\subsection{Numerical set-up and scale variation prescription}
\label{subsec:set-up}
In this section, we review the calculational set-up as well as the kinematical
constraints imposed to obtain the fiducial cross sections for Z+$\cjet$
production.  To select our final-state events, we focus on the forward region
with fiducial cuts mirroring those of the LHCb measurement~\cite{LHCb:2021stx}
at $\sqrt{s}=13$~TeV.

In particular, the following fiducial cuts for jets and charged leptons are
applied: $20~\GeV < p_{\rT,j} < 100~\GeV$, $2.2 < \eta_{j} < 4.2$, $p_{\rT,\ell}
> 20~\GeV$, $2.0 < y_{\ell} < 4.5$, $M_{\ell\bar{\ell}} \in[60,120]~\GeV$ and
$\Delta R(j,\ell) > 0.5$.  The jets are reconstructed with the anti-$k_T$
algorithm~\cite{Cacciari:2008gp} with $R=0.5$.
As discussed in Section~\ref{sec:calc}, the selection of $\cjets$ is performed
using the flavour dressing procedure described in~\cite{Gauld:2022lem}.
The algorithm proceeds in two stages with internal parameters that control the
overall flavour-tagging procedure: a flavour clustering stage that employs a
Soft-Drop-inspired criterion~\cite{Larkoski:2014wba} with parameters $z_{\rm
  cut}$, $R_{\rm cut}$, and $\beta$, followed by a flavour dressing stage based
on the flavour-$k_T$ distance measure~\cite{Banfi:2006hf} with a parameter
$\alpha$. In the present calculation, we set the parameters to their default
values~\cite{Gauld:2022lem}: $z_{\rm cut} = 0.1$, $R_{\rm cut} = 0.1$, $\beta =
2$, $\alpha = 2$.
In addition, events are only retained if the flavour-tagged $\cjet$ is the jet
carrying the largest transverse momentum of reconstructed jets passing the
selection cuts.

We provide predictions for proton-proton collisions at $\sqrt{s}=13$~TeV and use
the PDF4LHC21 Monte Carlo PDF set~\cite{PDF4LHCWorkingGroup:2022cjn}, with
$\alpha_s(M_\PZ) = 0.118$ and $n_f^{\rm max} = 5$, where both the PDF and
$\alpha_s$ values are accessed via LHAPDF~\cite{Buckley:2014ana}.
For the electroweak input parameters, the results are obtained in the
$G_{\mu}$-scheme, using a complex mass scheme for the unstable internal
particles, and we adopt the following values for the input parameters:
$M_{\PZ}^\mathrm{os} = 91.1876~\GeV$, $\Gamma_{\PZ}^\mathrm{os} = 2.4952~\GeV$,
$M_{\PW}^\mathrm{os} = 80.379~\GeV$, $\Gamma_{\PW}^\mathrm{os} = 2.085~\GeV,$
and $G_\mu = 1.1663787 \times 10^{-5}~\GeV^{-2}$.

For differential distributions, the impact of missing higher-order corrections
is assessed using the conventional 7-point scale variation prescription: the
values of factorisation ($\mu_F$) and renormalisation ($\mu_R$) scales are
varied independently by a factor of two around the central scale $\mu_0 \equiv
E_{\rT,\PZ}$, with the additional constraint that $\frac{1}{2} \leq \mu_F/\mu_R
\leq 2$.

When considering theoretical predictions for the ratio of distributions, we
estimate the uncertainties in an uncorrelated way between the numerator and
denominator i.e.\ by considering
\begin{equation} \label{eq:ratio_uncertainty}
  R^{c}_{j} (\mu_R,\mu_F;\mu'_R,\mu'_F) = \frac{\sigma^{\PZ+\cjet}(\mu_R,\mu_F)}{\sigma^{\PZ+\jet}(\mu'_R,\mu'_F)}\,,
\end{equation}
providing a total of 31-points when dropping the extreme variations in any pair
of scales.

\subsection{$Z+c$-jet distributions}
\label{subsec:dist1}

\begin{figure*}[t]
  \centering
  \begin{subfigure}[h]{.40\textwidth}
    \includegraphics[width=\textwidth]{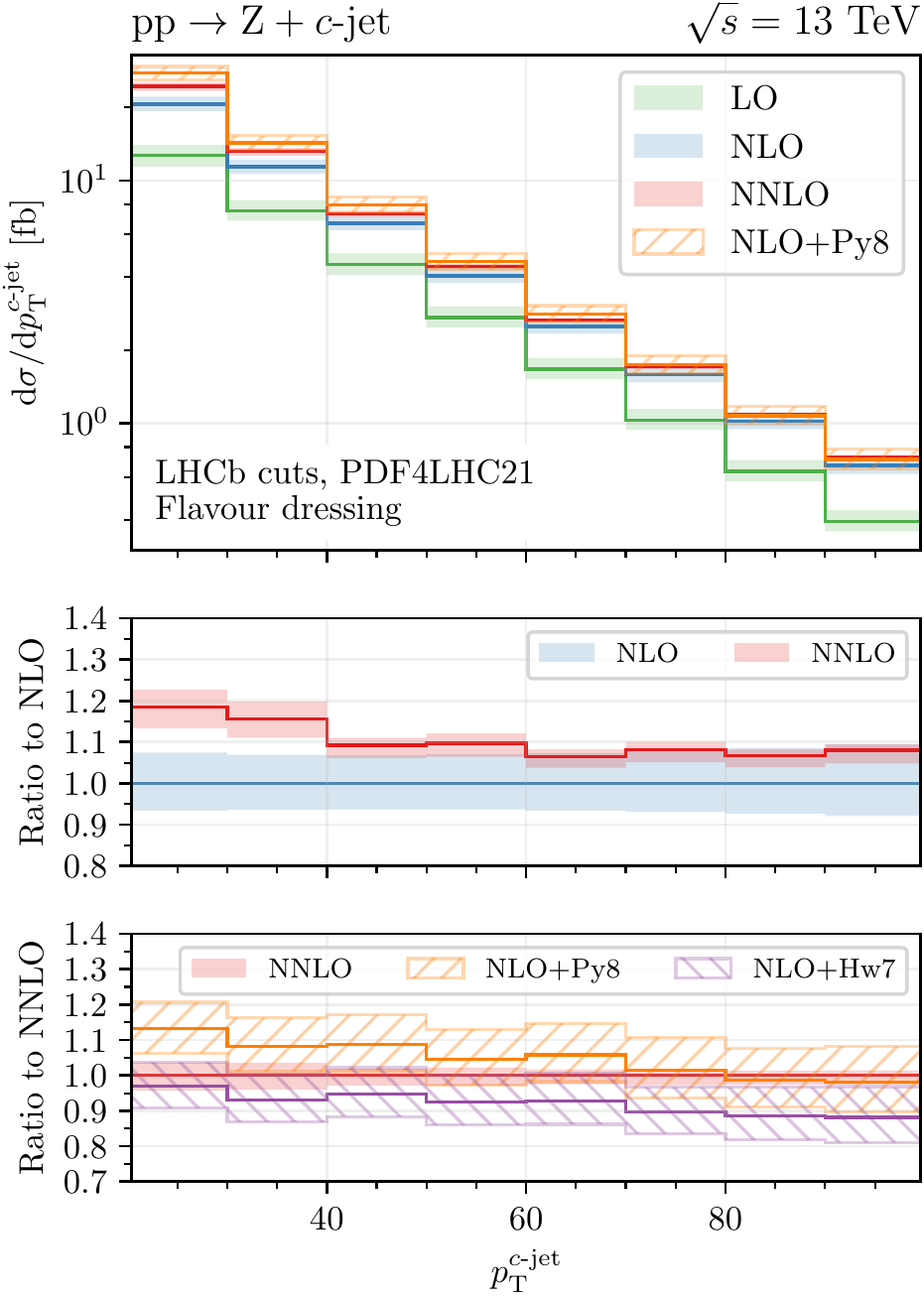}
    \subcaption{}
    \label{fig:zc-ptc-nocut}
  \end{subfigure}
  \hspace{1cm}
  \begin{subfigure}[h]{.40\textwidth}
    \includegraphics[width=\textwidth]{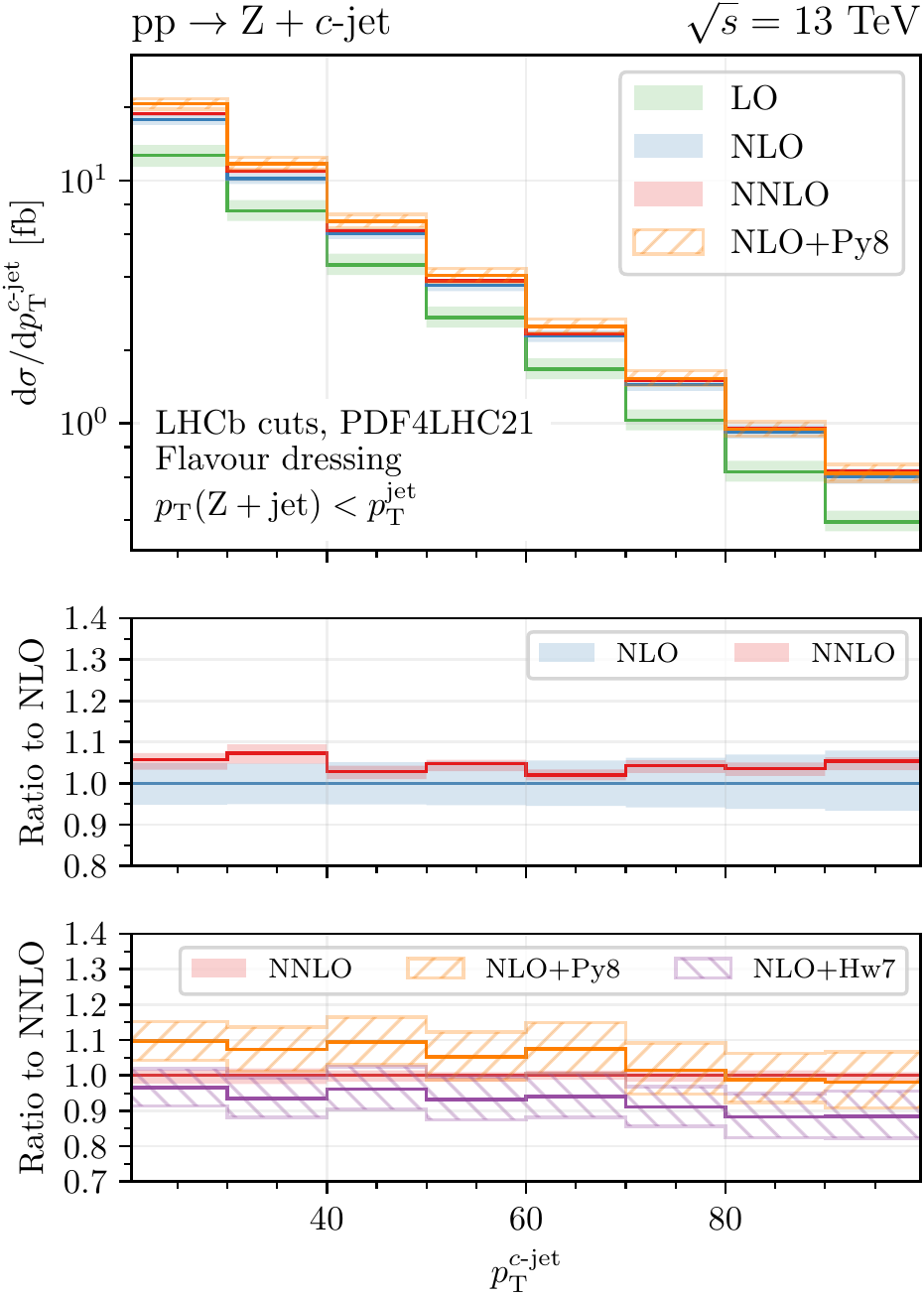}
    \subcaption{}
    \label{fig:zc-ptc-cut}    
  \end{subfigure}
  \caption{ Comparison of parton-level predictions for the leading flavoured jet
    transverse momentum $p_{\rT}^{\cjet}$ in the $\PZ+\cjet$ process: fixed-order
    predictions at \LO (green), \NLO (blue) and \NNLO (red); \NLOPS predictions
    with \PYTHIA (orange) or \HERWIG (purple) as parton showers.
    A dynamical cut on the transverse momentum of the $\PZ+\jet$ system is
    further applied in~(\subref{fig:zc-ptc-cut}).}
  \label{fig:zc-ptc}
\end{figure*}

\begin{figure*}[t]
  \centering
  \begin{subfigure}[h]{.40\textwidth}
    \includegraphics[width=\textwidth]{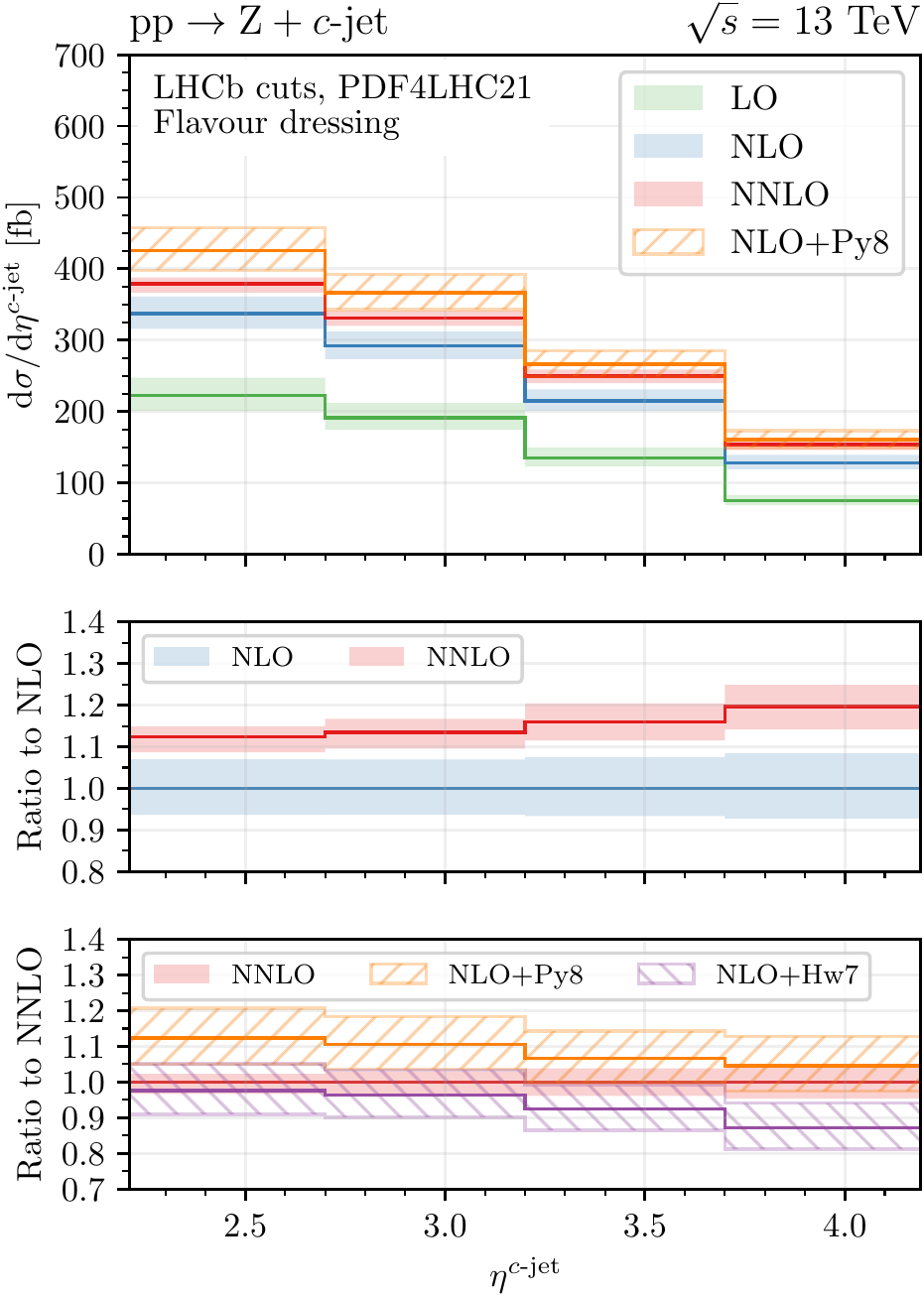}
    \subcaption{}
    \label{fig:zc-etac-nocut}
  \end{subfigure}
  \hspace{1cm}
  \begin{subfigure}[h]{.40\textwidth}
    \includegraphics[width=\textwidth]{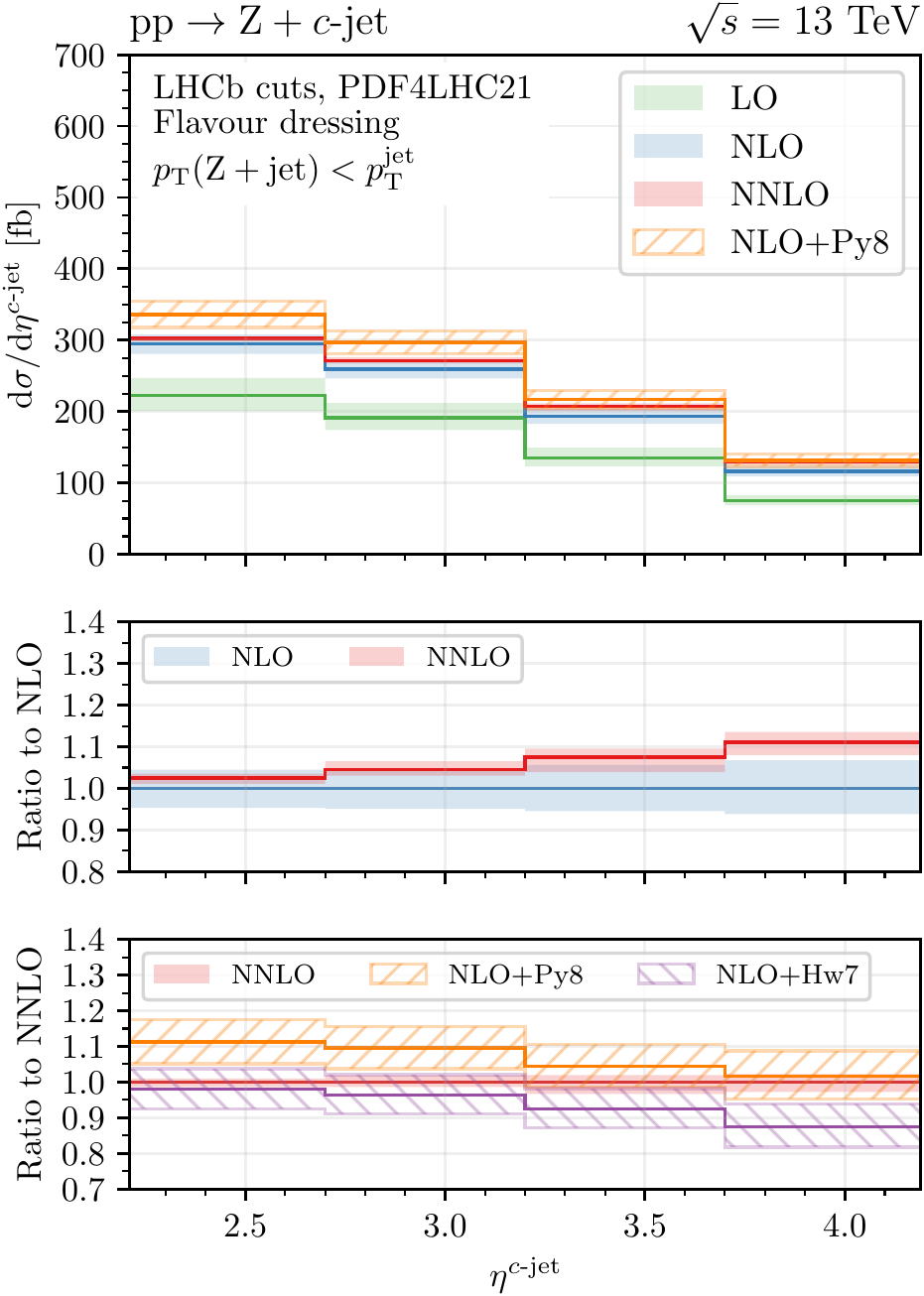}
    \subcaption{}
    \label{fig:zc-etac-cut}    
  \end{subfigure}  
  \caption{Comparison of parton-level predictions for the leading flavoured jet
   pseudo-rapidity $\eta^{\cjet}$ in the $\PZ+\cjet$ process: fixed-order
    predictions at \LO (green), \NLO (blue) and \NNLO (red); \NLOPS predictions
    with \PYTHIA (orange) or \HERWIG (purple) as parton showers.
    A dynamical cut on the transverse momentum of the $\PZ+\jet$ system is
    further applied in~(\subref{fig:zc-etac-cut}). }
  \label{fig:zc-etac}
\end{figure*}

\begin{figure*}[t]
  \centering
  \begin{subfigure}[h]{.40\textwidth}
    \includegraphics[width=\textwidth]{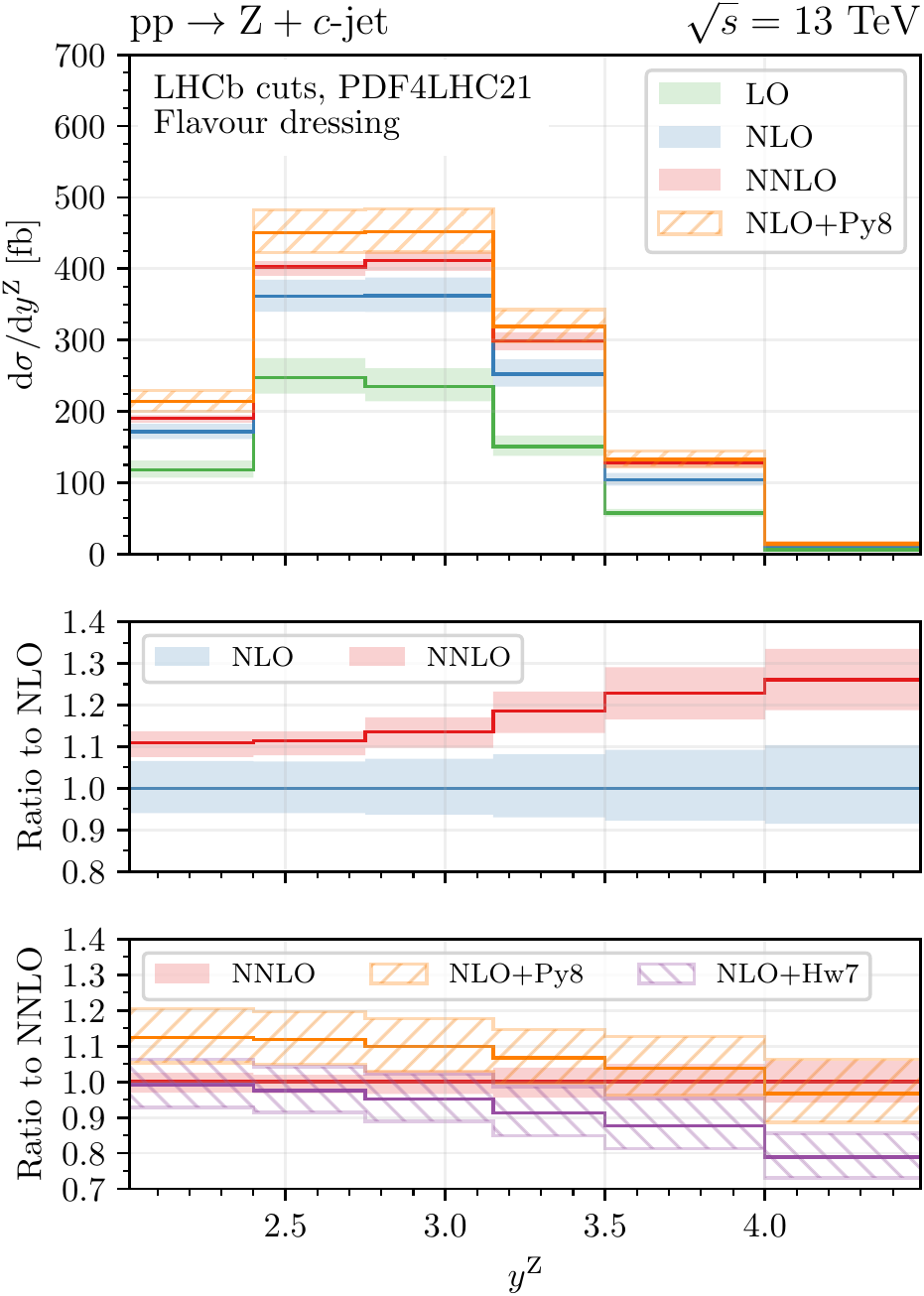}
    \subcaption{}
    \label{fig:zc-yz-nocut}
  \end{subfigure}
  \hspace{1cm}
  \begin{subfigure}[h]{.40\textwidth}
     \includegraphics[width=\textwidth]{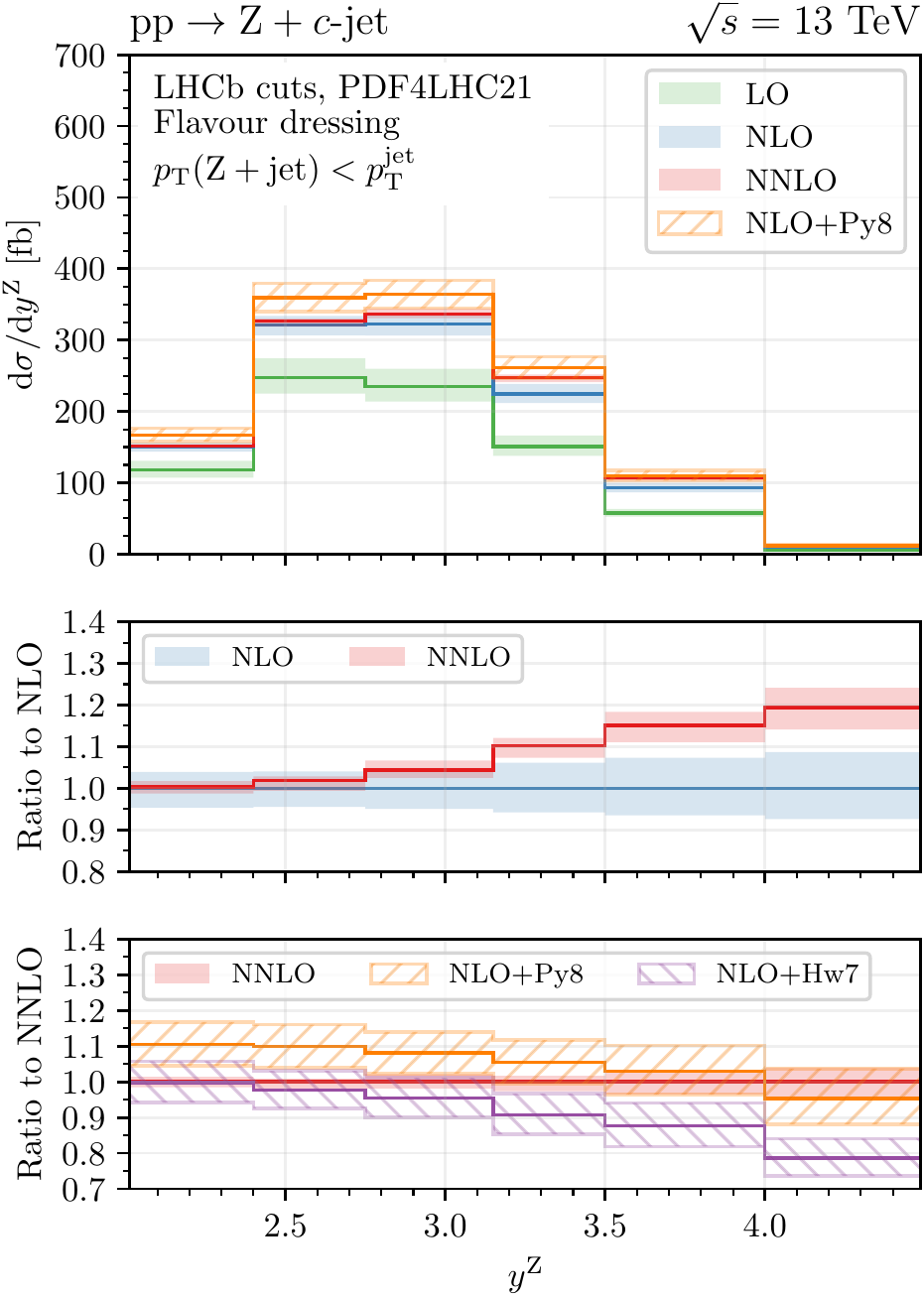}
    \subcaption{}
    \label{fig:zc-yz-cut}    
  \end{subfigure}
  \caption{Comparison of parton-level predictions for the rapidity distribution
    of the lepton pair $y^Z$ in the $\PZ+\cjet$ process: fixed-order predictions
    at \LO (green), \NLO (blue) and \NNLO (red); \NLOPS predictions with \PYTHIA
    (orange) or \HERWIG (purple) as parton showers. A dynamical cut on the
    transverse momentum of the $\PZ+\jet$ system is further applied
    in~(\subref{fig:zc-yz-cut}).}
 \label{fig:zc-yz}
\end{figure*}

We here present results for the $\PZ+\cjet$ process at $\sqrt{s}=13$~TeV and
choose to focus on the following observables: the leading flavoured jet
transverse momentum $p_{\rT}^{\cjet}$ (Figure~\ref{fig:zc-ptc}), the leading
flavoured jet pseudorapidity $\eta^{\cjet}$ (Figure~\ref{fig:zc-etac}), and the
rapidity of the $\PZ$-boson $y^{\PZ}$, reconstructed from the two final-state
opposite-charge leptons (Figure~\ref{fig:zc-yz}). Besides presenting results for
the LHCb kinematical set-up as indicated in Section~\ref{subsec:set-up}, we also
explore the impact of the introduction of a cut on the transverse momentum of
the $\PZ+\jet$ system:
\begin{equation}\label{eq:cut-ptzj}
  p_{\rT}(\PZ+\jet) < p_{\rT}^{\jet}\,,
\end{equation}
with the leading jet in the acceptance region.
The theoretical motivation behind this cut is to discard those contributions
where the flavoured jet is not the jet with the largest transverse momentum in
the event, i.e.\ cases where the hardest jet was disregarded because it fell
outside of the LHCb acceptance.
At Born level, the $p_{\rT}$ of the $\PZ+\jet$ system vanishes, hence the cut
in~\eqref{eq:cut-ptzj} limits the hard QCD radiation outside the LHCb acceptance
in a dynamical way.

For each of the figures presented in the remainder of this Section, the left
sides present results without the additional cut of~\eqref{eq:cut-ptzj}, whereas
the right sides present results where the additional cut has been imposed, such
that the impact of the cut can be seen comparing the left and right hand sides
of the figures.
To highlight the size and shape of the fixed-order results at
each perturbative order, and to best compare fixed-order with
and without matching to PS results, all figures illustrating the results in this
Section are composed of three panels: the top-panel shows the absolute
predictions at fixed-order (LO, NLO, NNLO) and for NLO+PS where PS is modelled by
\PYTHIA; the middle panel shows the ratio of NNLO to the fixed-order NLO result
while the lower panel shows the ratio to NNLO of the NLO+PS results where PS is
modeled by either \PYTHIA or \HERWIG.  As noted in Section~\ref{sec:calc},
fixed-order predictions labelled as NLO and NNLO in all the figures include
QCD corrections up to ${\cal O}(\alpha_{s}^{2})$ (NLO) and
${\cal O}(\alpha_{s}^{3})$ (NNLO) obtained with massless c-quarks in the
computations, and both additionally include the exact charm-quark mass corrections up to ${\cal O}(\alpha_{s}^{2})$.

The theory predictions which \emph{do not} include the additional kinematic
constraint of Eq.~\eqref{eq:cut-ptzj} are shown in
Figs.~\ref{fig:zc-ptc-nocut},~\ref{fig:zc-etac-nocut},
and~\ref{fig:zc-yz-nocut}.  The NNLO corrections are observed to be of the order
(10--20)\% as compared to the NLO prediction, and typically outside of the scale
variation band of the NLO result.
In particular, the shape of the distribution is modified in the
large-$\eta^{\cjet}$ and large-$y^{\PZ}$ bins in Fig.~\ref{fig:zc-etac-nocut} in
Fig.~\ref{fig:zc-yz-nocut} respectively, and in the small-$p_{\rT}^{\cjet}$
region in Fig.~\ref{fig:zc-ptc-nocut}, featuring a larger cross section at NNLO.
Those features are made apparent in the second panel of those figures.
In the third panel, the NNLO and both NLO+PS predictions are shown normalised to
the central NNLO result.  It is found that the NNLO result always lies between
the two different NLO+PS results.  We observe that the NNLO result is more
consistent with the NLO+\HERWIG prediction for lower values of
$p_{\rT}^{\cjet}$, $\eta^{\cjet}$, and $y^{\PZ}$, and instead agrees better with
NLO+\PYTHIA predictions at larger values.
Overall, the NLO+\HERWIG prediction seems to be rather similar to the NLO
prediction, and the angular-ordered parton shower does not appear to impact
flavour-sensitive observables in a significant way.
Instead, the NLO+\PYTHIA prediction does seem to capture some of the NNLO higher
order corrections: at large-$p_{\rT}^{\cjet}$,$\eta^{\cjet}$ and $y^{\PZ}$
values it tends to reproduce the shape of the fixed-order NNLO corrections.

The theory predictions that \emph{do} include the additional kinematic
constraint of Eq.~\eqref{eq:cut-ptzj} are shown in
Figs.~\ref{fig:zc-ptc-cut},~\ref{fig:zc-etac-cut}, and~\ref{fig:zc-yz-cut}.  We
observe that the constraint leads to a slight reduction of the fiducial cross
section and produces no significant change in the shape of the distributions
with the exception of the low-$p_{\rT}^{\cjet}$ region.
The LO, NLO, and NNLO results display an improved mutual compatibility across
all considered observables, indicating a better perturbative convergence in the
presence of this kinematic constraint.
Qualitatively, the comparison between the NNLO and NLO+PS results in the third
panel of these figures is similar to that of the case without the kinematic
constraint which was already discussed.

Overall, for all considered set-ups, we find that the NNLO corrections bring a
new level of precision for the considered $\PZ+\cjet$ observables.  As compared
to the corresponding NLO(+PS) results, the uncertainties of the NNLO predictions
are reduced by a factor of two or more.
It is also reassuring that the NNLO corrections lead to predictions that tend to
lie between the two different NLO+PS predictions---the latter differ in the
treatment of ${\cal O}(\alpha_{s}^{3})$ terms, which begin at the NNLO level.
As the perturbative convergence of the predictions appears to be improved by
applying the kinematic constraint of Eq.~\eqref{eq:cut-ptzj}, with only a small
reduction in the cross section, it seems to be well motivated when directly
considering $\PZ+\cjet$ observables.

\subsection{The \texorpdfstring{$\sigma(\PZ+\jet)$}{Z+jet} cross-section and the ratio \texorpdfstring{$R^{c}_{j}$}{Rcj}}
\label{subsec:ratio}

In Section~\ref{subsec:dist1} we have presented IRC safe predictions for the
rates of $\PZ+\cjet$ production within the LHCb fiducial region.  Instead, in
this subsection we will consider the $\PZ+\jet$ process (i.e.\ the flavour
inclusive one) and then subsequently the ratio observable $R^{c}_{j} =
\sigma(\PZ+\cjet)/\sigma(\PZ+\jet)$.  As the experimental measurement of
$R^{c}_{j}$ is performed differentially in the rapidity of the $\PZ$-boson
$y^{\PZ}$~\cite{LHCb:2021stx}, our theory predictions will also focus on this
same quantity.
We again note that the we do not perform any comparison to the available data
for reasons of consistency, as detailed in Appendix~\ref{sec:irc}
and~\ref{sec:mpi}.

\begin{figure*}[t]
  \centering
  \begin{subfigure}[h]{.40\textwidth}
    \includegraphics[width=\textwidth]{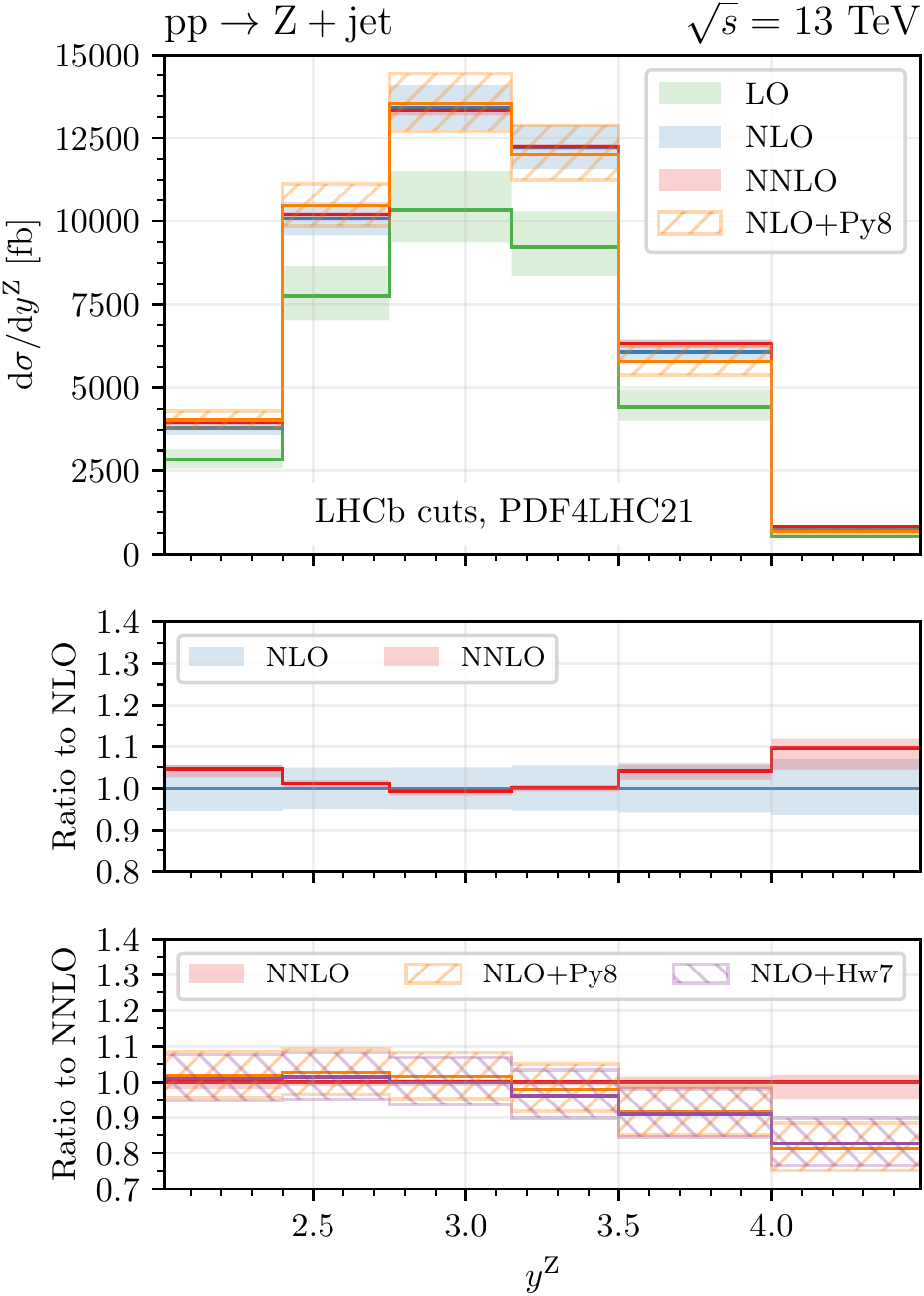}
    \subcaption{}
    \label{fig:zj-yz-nocut}
  \end{subfigure}
  \hspace{1cm}
  \begin{subfigure}[h]{.40\textwidth}
     \includegraphics[width=\textwidth]{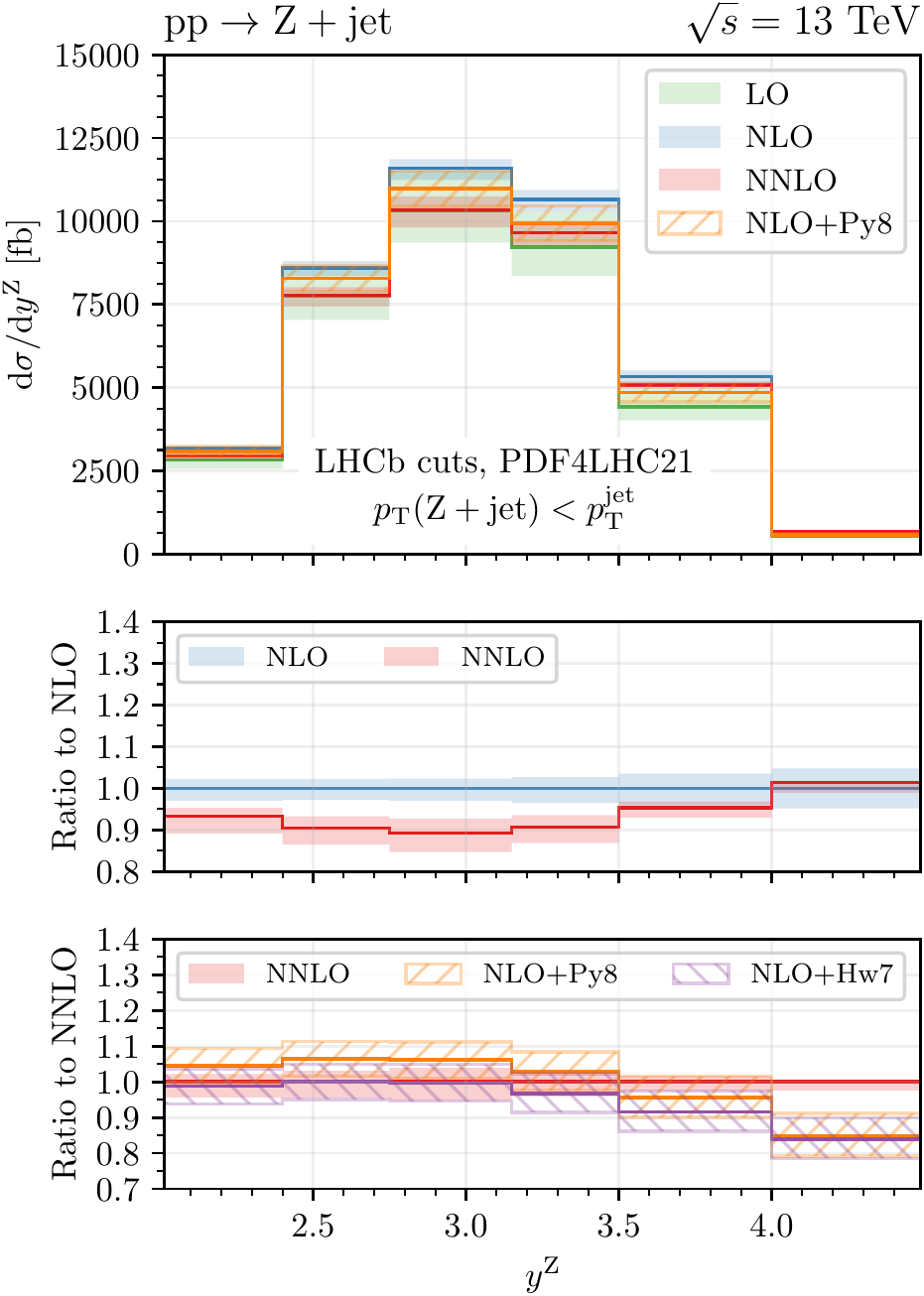}
    \subcaption{}
    \label{fig:zj-yz-cut}    
  \end{subfigure}
  \caption{Comparison of parton-level predictions for the rapidity distribution of the lepton pair $y^Z$ for the unflavoured process $\PZ+\jet$: fixed-order predictions at \LO (green), \NLO (blue) and \NNLO (red); \NLOPS predictions
    with \PYTHIA (orange) or \HERWIG (purple) as parton showers.
A dynamical cut on the transverse momentum of the $\PZ+\jet$ system is
further applied in~(\subref{fig:zj-yz-cut}).}   
  \label{fig:zj-yz}
\end{figure*}

The theory predictions for $\PZ+\jet$ production are presented in
Fig.~\ref{fig:zj-yz}, with the same structure for the plots as shown in
Section~\ref{subsec:dist1}, i.e.\ with the ``no-cut'' and ``with-cut'' cases
shown on the left and right parts of the figure respectively, and with three
panels for each sub-figure.
As compared to the $\PZ+\cjet$ predictions, heavy-flavour mass corrections have
not been included for these predictions.  While this could be achieved following
the procedure outlined in~\cite{Gauld:2021zmq}, the numerical impact of such
corrections for the flavour inclusive process is negligible (sub-percent).

The theory predictions without the extra kinematic cut of
Eq.~\eqref{eq:cut-ptzj} are shown in Fig.~\ref{fig:zj-yz-nocut}.  The NNLO
result is observed to be contained within the scale variation band of the NLO
one, except in the large-$y^{\PZ}$ region, where it also features a different
behaviour compared to the \NLOPS results.
We also find very good agreement between the two \NLOPS results with \PYTHIA and
\HERWIG.

The impact of applying the additional kinematic cut is shown in
Fig.~\ref{fig:zj-yz-cut}.  The cut leads to negative NNLO corrections in the
(relatively) low $y^{\PZ}$ region, resulting in a NNLO prediction lying outside
the scale variation band of the NLO result, except in the large-$y^{\PZ}$
region.
Examination of the upper panel of Fig~\ref{fig:zj-yz-cut} shows that the cut has
the effect of moving all the curves closer to the LO result, with positive NLO
corrections and negative NNLO corrections.
The inclusion of this cut thus appears to degrade the perturbative stability for
the flavour-inclusive set-up.

\begin{figure*}[t]
  \centering
  \begin{subfigure}[h]{.40\textwidth}
    \includegraphics[width=\textwidth]{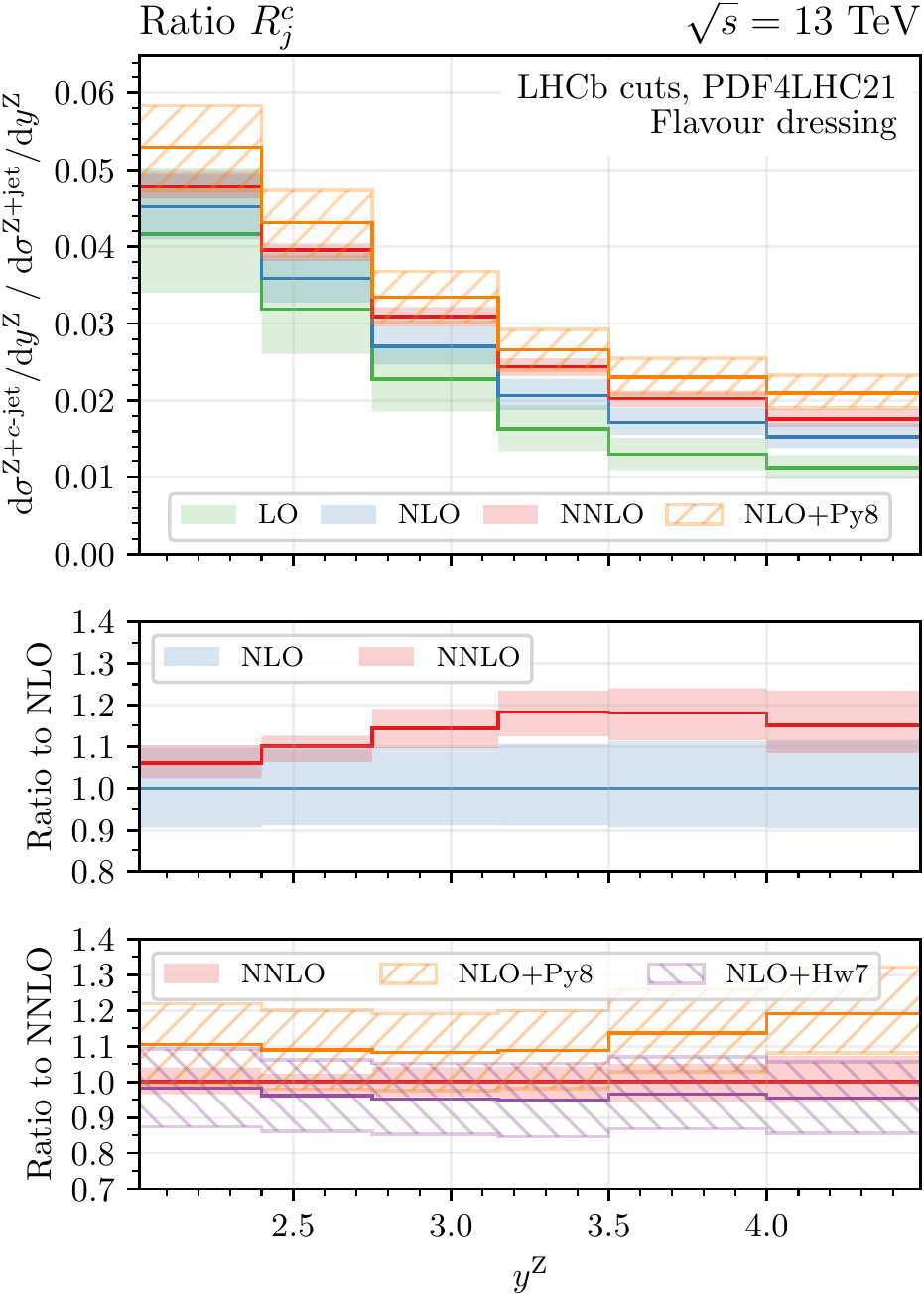}
    \subcaption{}
    \label{fig:ratio-nocut}
  \end{subfigure}
  \hspace{1cm}
  \begin{subfigure}[h]{.40\textwidth}
    \includegraphics[width=\textwidth]{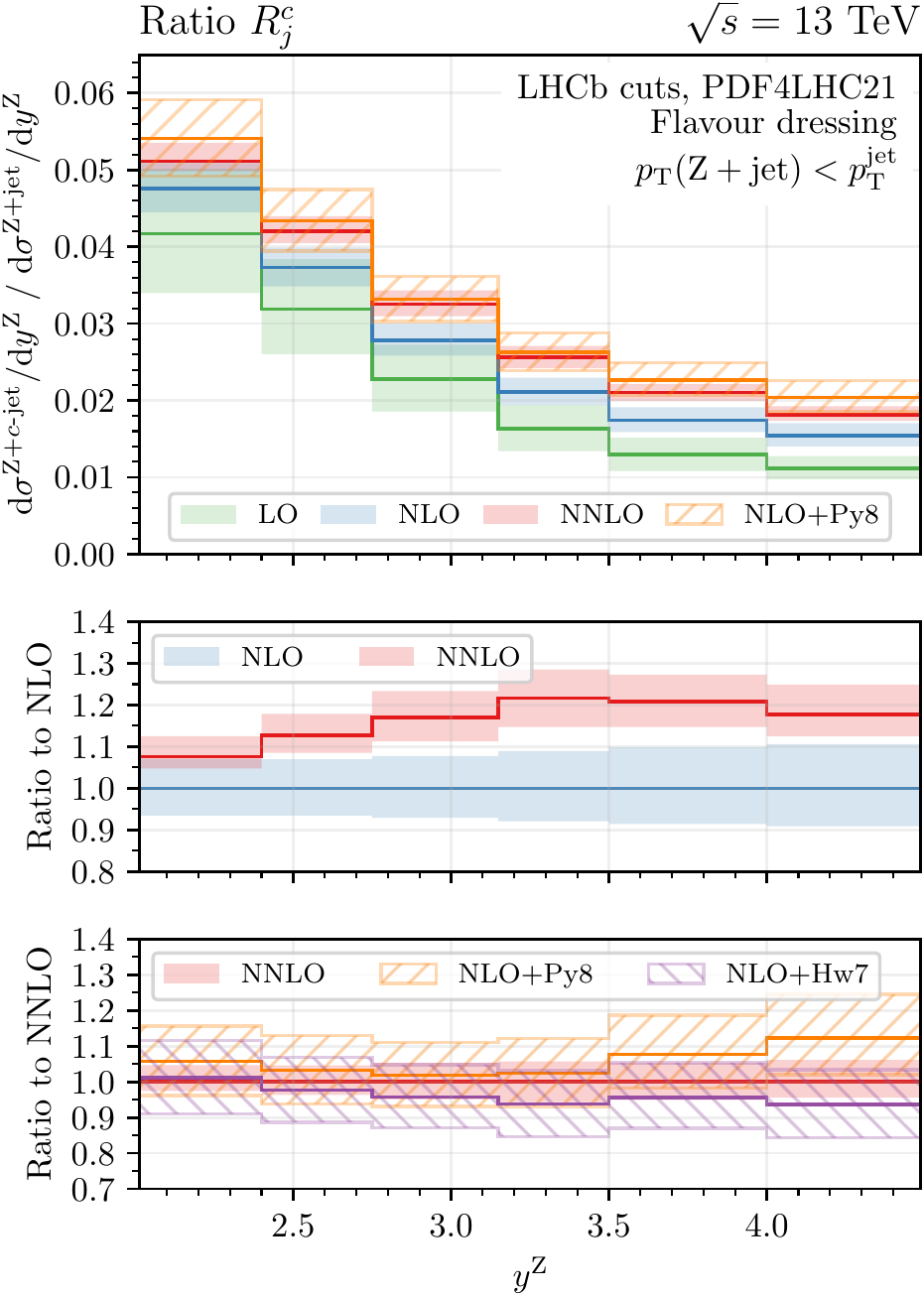}
    \subcaption{}
    \label{fig:ratio-cut}    
  \end{subfigure}  
  \caption{ Comparison of parton-level predictions for the ratio $R^{c}_{j} =
    \sigma^{\PZ+\cjet}/\sigma^{\PZ+\jet}$ differential in the rapidity $y^{\PZ}$
    of the system of the final-state leptons: fixed-order predictions at \LO
    (green), \NLO (blue) and \NNLO (red); \NLOPS predictions with \PYTHIA
    (orange) or \HERWIG (purple) as parton showers.  A dynamical cut on the
    transverse momentum of the $\PZ+\jet$ system is further applied
    in~(\subref{fig:ratio-cut}).}
\label{fig:ratio}
\end{figure*}

We now consider the ratio observable $R^{c}_{j}$, differential in $y^{\PZ}$.
The result is constructed using the same inputs which lead to the distributions
for $\PZ+\cjet$ results in Fig.~\ref{fig:zc-yz} and $\PZ+\jet$ in
Fig.~\ref{fig:zj-yz}, but including the uncorrelated uncertainty prescription
defined in Eq.~\eqref{eq:ratio_uncertainty}.
The predictions for $R^{c}_{j}$ are displayed in Fig.~\ref{fig:ratio}, again the
``no-cut'' and ``with-cut'' cases are shown on the left and right sides of the
figure respectively.  Focussing first on the fixed-order results, the NNLO
corrections are observed to be positive and of the order (10--20)\% with the
largest value observed at large-$y^{\PZ}$ values.  That behaviour is observed
for both the ``no-cut'' (left) and ``with-cut'' (right) cases.  Overall, the
inclusion of the cut on the $\PZ+\jet$ system does not significantly impact the
perturbative behaviour of the fixed-order prediction for the ratio.
This is a consequence of the fact that the reduction of uncertainties in the
numerator for $\PZ+\cjet$ is then compensated by increased uncertainties in the
denominator, when the cut is applied.
However, by inspecting the lowest panels of Fig.~\ref{fig:ratio}, we observe
slightly better agreement between NNLO and the two NLO+PS predictions when the
cut has been applied.

Finally, as a result of the uncorrelated prescription for uncertainties, we note
that the relative theory uncertainties for the NNLO predictions of the ratio are
increased as compared to the individual predictions for $\PZ+\cjet$ and
$\PZ+\jet$.  The sensitivity to the input PDFs is also typically reduced for
such an observable due to correlations between PDF-dependence of the numerator
and denominator.
With the aim of reducing the theory uncertainties due to missing higher
corrections, one could therefore consider to include absolute $\PZ + \cjet$
cross-section data rather than that for the ratio $R^c_j$ in a collinear PDF
fit. Given however that several experimental uncertainties are correlated
between numerator and denominator (and therefore cancel in the ratio), and that
a treatment of the MPI contribution to the observable should also be considered,
overall it is not clear which observable is the most sensitive in constraining
PDFs.

\section{Conclusions}
\label{sec:concl}

In this paper, we have studied the associated production of a $\PZ$-boson with a
charm-jet at the LHC at 13\,TeV in the forward region.
We computed NNLO predictions at $\mathcal{O}(\alpha_s^3)$ for a set of
differential observables related to the $\PZ+\cjet$ process using the flavour
dressing procedure to define charm-tagged jets.  NNLO corrections are found to
be at the level of (10--20)\%; they can impact the shapes of distributions with
the high-$y^{\PZ}$ and low-$p_{\rT}^{\cjet}$ regions receiving enhanced
corrections.  The residual uncertainties as estimated through scale variations
are typically $\pm5$\% or smaller, a factor two reduction compared to the
respective NLO uncertainty estimate.  Additionally, comparisons to two different
\NLOPS predictions based on the \HERWIG (angular ordering) and \PYTHIA ($p_\rT$
ordering) showers have been performed.  The two predictions can differ by up to
10\% but remain mutually compatible within their respective uncertainties that
are NLO-like and thus at the level of $\pm10$\%.  The NNLO distributions are
found to lie in between the two \NLOPS predictions, hinting to an insensitivity
to missing higher-order effects as modelled by the showers.  Moreover, we have
found that a theory-inspired constraint on the transverse momentum of the
$\PZ+\jet$ system improves the perturbative convergence of all considered
distributions.

We further considered the ratio $R^c_j = \sigma^{\PZ+\cjet}/\sigma^{\PZ+\jet}$,
which has been measured by the LHCb experiment.  In this context, the usage of
the flavour dressing procedure ensures the same kinematic reconstruction of jets
entering the numerator and denominator, thus allowing for a faithful theoretical
definition.  The pattern of higher-order corrections mimics those of the
$\PZ+\cjet$ process with enhanced NNLO corrections of up to 20\% in the
high-$y^{\PZ}$ region, albeit with larger uncertainties due to the de-correlated
scale prescription for the ratio.

A direct comparison to the available LHCb data was not performed due to IRC
unsafety issues and an unexpectedly large contamination from MPI; both of which
are discussed in the Appendices.  In the future, a fair comparison between
experimental measurements and theory predictions will require a detailed study
of the experimental feasibility of the flavour dressing algorithm (or other IRC
safe variants).
In the case a direct application of an IRC-safe flavour definition is
prohibitively challenging, it would be highly desirable for experimental
measurements to carry out an unfolding to a IRC-safe definition of jet flavour.
Only a joint effort of both communities, theory and experimental, will enable to
exploit in the best way the huge amount of data that LHC will provide us in the
next decades, better enabling the use flavour signatures as a powerful window
into short-distance interactions from $\GeV$ to $\TeV$ energy scales.

\begin{acknowledgements}
We thank Marco Zaro for his help with the usage of MadGraph5 aMC@NLO and
Christian Preuss for discussions about the treatment of heavy quarks in Pythia8.
RG and AH thank Philip Ilten for several discussions related to the LHCb, and
more generally on the experimental definition of flavoured jets.  GS thanks the
Institute for Theoretical Physics at ETH for hospitality during the course of
this work.  This research was supported in part by the UK Science and Technology
Facilities Council under contract ST/T001011/1 and by the Swiss National Science
Foundation (SNF) under contracts 200021-197130 and 200020-204200.  Numerical
simulations were facilitated by the High Performance Computing group at ETH
Z\"urich and the Swiss National Supercomputing Centre (CSCS) under project ID
ETH5f.
\end{acknowledgements}

\begin{appendices}
    
\section{IRC safety} \label{sec:irc}

According to Ref.~\cite{LHCb:2021stx}, as part of the LHCb measurement, charm
jets are defined in the following way.
First, jets are reconstructed using the anti-$k_T$
algorithm~\cite{Cacciari:2008gp} with $R=0.5$, and the leading jet passing the
fiducial selection is considered (see also Section~\ref{subsec:set-up} for the
definition of the fiducial selection).
From this, the leading jet is considered to be a charm jet (at truth/unfolded
level) if it additionally satisfies the criterion: $p_{\rT,c{~\rm hadron}} >
5~\GeV$, and $\Delta R(j,c{~\rm hadron}) < 0.5$.
That is to say, the jet is considered to be charm tagged if there is \textbf{at
  least} one $c{~\rm hadron}$ satisfying these selections.

Such an experimental definition of jet flavour is collinear unsafe.
This is not an LHCb specific issue (or even particular to this measurement), but
is also relevant in one way or another for the definitions of jet flavour
commonly adopted by LHC
collaborations~\cite{LHCb:2015tna,ATLAS:2015thz,CMS:2017wtu,LHCb:2021dlw}.
We briefly explain why the observable definition taken in
Ref.~\cite{LHCb:2021stx} is IRC unsafe, then discuss the implications of this
for data interpretation.

\begin{figure*}[t]
  \centering
  \begin{subfigure}[b]{.18\textwidth}
    \centering    
    \includegraphics[width=\textwidth]{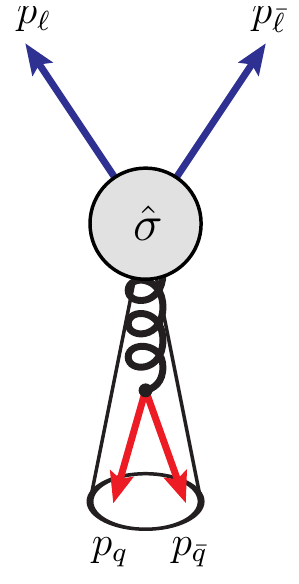}
    \subcaption{}
    \label{fig:IRC1}
  \end{subfigure}
  \,\,\,\,\,\,\,\,\,\,\,\,
  \begin{subfigure}[b]{.18\textwidth}
    \centering    
    \includegraphics[width=\textwidth]{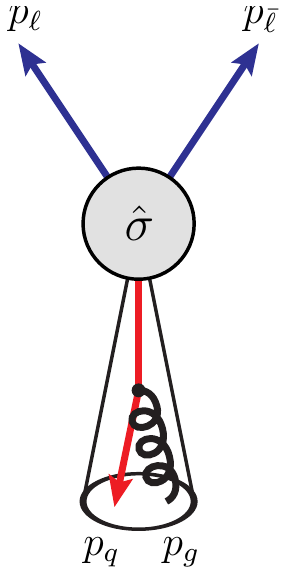}
    \subcaption{}
    \label{fig:IRC2}
  \end{subfigure}
  \,\,\,\,\,\,\,\,\,\,\,\,
  \begin{subfigure}[b]{.22\textwidth}
    \centering
    \includegraphics[width=\textwidth]{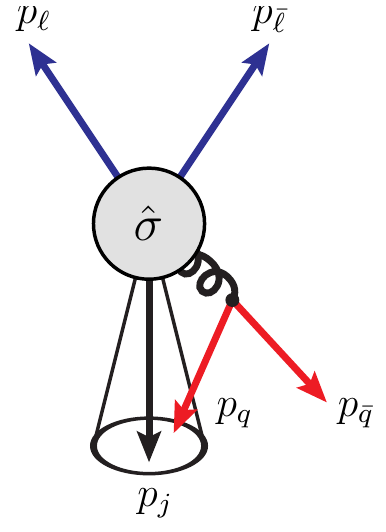}
    \subcaption{}
    \label{fig:IRC3}
  \end{subfigure}
  \caption{Examples of configurations which lead to an IRC sensitivity:
    (\subref{fig:IRC1}) a double (even) tagged jet; (\subref{fig:IRC2}) a tag
    with a $p_{\rT,q}^{\rm min}$ requirement; (\subref{fig:IRC3}) a jet tag with
    a soft sensitivity (in the absence of a $p_{\rT,q}^{\rm min}$ requirement).}
  \label{fig:IRCconfigurations}
\end{figure*}

In Fig.~\ref{fig:IRCconfigurations} we depict three specific kinematic
configurations which lead to different sources of IRC sensitivity (i.e.\ those
which are IRC unsafe). In each case, the configurations correspond to those
encountered in a parton-level fixed-order prediction for the $\PZ+\cjet$
process. The flavoured quarks are depicted as red lines, i.e.\ representing
charm quarks for the process under consideration here.
\begin{enumerate} [label=(\alph*)]
\item The first configuration depicts the production of the lepton pair,
  recoiling against a hard $q\bar q$ pair produced in a collinear configuration
  (i.e.\ at least one, or both, of the quarks are hard but
  $p_q \cdot p_{\bar{q}} \to 0$). When the anti-$k_{\rT}$ algorithm is applied,
  the $q$ and $\bar q$ are reconstructed inside the same jet. According to the
  prescription to assign jet flavour, as this jet contains \textbf{at least} one
  quark it will be assigned a quark flavour tag (e.g. charm for $q = c$).
	This introduces a collinear sensitivity as it is distinguished from the case
  where the hard gluon does not split into a collinear $q\bar q$ pair. In that
  case, the jet (composed of a single hard gluon) clearly carries zero quark
  flavour. This could be overcome by accounting for the total quark flavour in
  the jet, such as assigning quantum numbers $q(\bar q)= +(-)1$ and summing them
  to obtain the net flavour (alternatively one can consider flavoured jets as
  those with an overall odd number of $q$ and $\bar q$).
\item The second configuration depicts the production of the lepton pair,
  recoiling against a hard $qg$ pair in a collinear configuration (i.e.\ at least
  one, or both, of the quark and gluon is hard but $p_q \cdot p_{g} \to
  0$). Again, when the anti-$k_{\rT}$ algorithm is applied, both $q$ and $g$ are
  reconstructed inside the same jet.
	As the tagging prescription requires the presence
  of a $c$-hadron with $p_{\rT,c} > 5~\GeV$, it is possible that the outgoing
  quark does not satisfy this criterion (depending on the momentum sharing with
  the gluon). This introduces a collinear sensitivity as the $p_{\rT,c}$
  requirement may distinguish from the case where the hard initial quark does
  not split to the collinear $qg$ pair.	
\item The collinear sensitivity discussed above can be overcome by removing the
  $p_{\rT,c}^{\rm min}$ requirement. However, this would introduce a new problem
  which is depicted in the third configuration. In this case, the culprit is a
  soft gluon which subsequently splits to a $q\bar q$ pair at wide angles. It is
  possible that one of the quarks ($p_q$ in the figure) is produced close in
  $\Delta R$ to a hard parton ($p_j$), i.e.\ $\Delta R(j,q) < 0.5$. This would
  introduces a soft sensitivity as the flavour of the jet would be altered by
  the presence of the soft quark.
\end{enumerate}

From a purely experimental point of view, it is clear that the current
definition of tagging heavy-flavour jets is a sensible one.
Identifying those jets with multiple tags (as oppose to at least one) requires
to more carefully account for the experimental (in)efficiency and mistag rates.
It is also extremely to difficult distinguish between the signature of one or two
collinear heavy-flavour objects (e.g. a bunch of displaced tracks appearing to
originate from a single displaced vertex).  Furthermore, removing the
$p_{\rT,c}^{\rm min}$ would mean accounting for a region where it is
experimentally challenging to identify displaced vertices.
However, this choice has serious ramifications for the theory predictions, and
importantly for the interpretation of the data.

Theoretical predictions of charm jet observables which are not IRC safe are
logarithmically sensitive to the mass of the charm quark $m_c$.
The corresponding fixed-order predictions for such observables therefore include
corrections which depend logarithmically on the charm quark mass. If the
observable/process under consideration involves energy scales which are large
compared to the quark mass (e.g. the transverse energy/momentum of a jet or a
boson), the logarithmic corrections become large (due to the separation of
scales) and thus limit the theory precision/perturbative stability.
The $m_c\to0$ limit of such predictions is not well defined (it is divergent),
meaning that a calculation based on massless quarks of such observables does not
exist.
The implications of this are that fixed-order predictions must be performed in a
scheme where the charm quark is massive, i.e.\ in a fixed-flavour-number scheme
with $n_f^{\rm max} = 3$, where mass factorisation is not performed for the
charm quark, and it is decoupled from the running of
$\alphas$.
Note that the perturbative charm-quark PDF does still exist in the massive scheme (where a
logarithmic sensitivity to the charm-quark mass exists, see for
example~\cite{Laenen:1992zk,Riemersma:1994hv,Harris:1995tu,Buza:1996wv,Laenen:1998kp,Kawamura:2012cr}). 
Practically, it is generated numerically after integration over the phase-space of the massive quark during the calculation.

The requirement that a fixed-order prediction must be carried out with a massive calculation 
is problematic for observables which are designed to be sensitive to the nature of the charm quark PDF.
Such observables contain a logarithmic sensitivity on the charm-quark mass as a
result of the IRC unsafe configurations which were highlighted above (which
limit the theory precision/predictability), at the same time the observables are
(by design) directly sensitive to the large logarithmic corrections associated
to the perturbative charm quark PDF which have not been resummed, and the
non-perturbative component of the charm quark PDF (which is aiming to be probed)
does not exist.

Instead, the definition of jet flavour used in this work, given its
IRC safety, removes all logarithmic sensitivity on the charm quark mass which
results from the jet flavour assignment procedure.
This also allows for the application of a massless calculation, where mass factorisation for the 
charm quark is performed and where remaining logarithmic sensitivity to the quark mass is resummed.

\section{Multiple Particle Interactions} \label{sec:mpi}

During the high-energy scattering of two protons, there is a probability for
multiple hard-interactions to occur (i.e.\ more than one).

\begin{figure*}[t]
  \centering
  \begin{subfigure}[h]{.30\textwidth}
    \includegraphics[width=\textwidth]{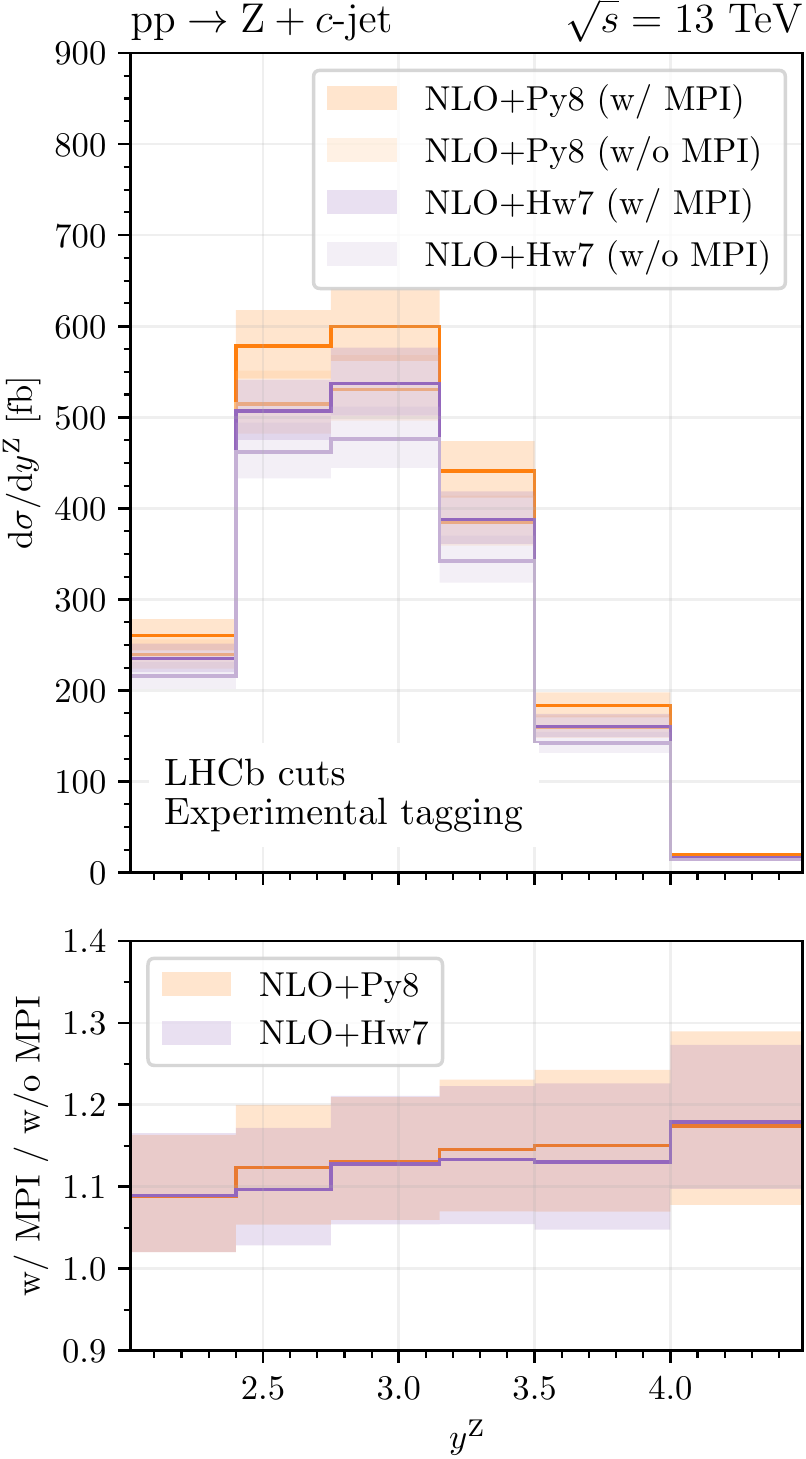}
    \subcaption{}
    \label{fig:MPI_zcjet}
  \end{subfigure}
  \,\,
  \begin{subfigure}[h]{.30\textwidth}
    \includegraphics[width=\textwidth]{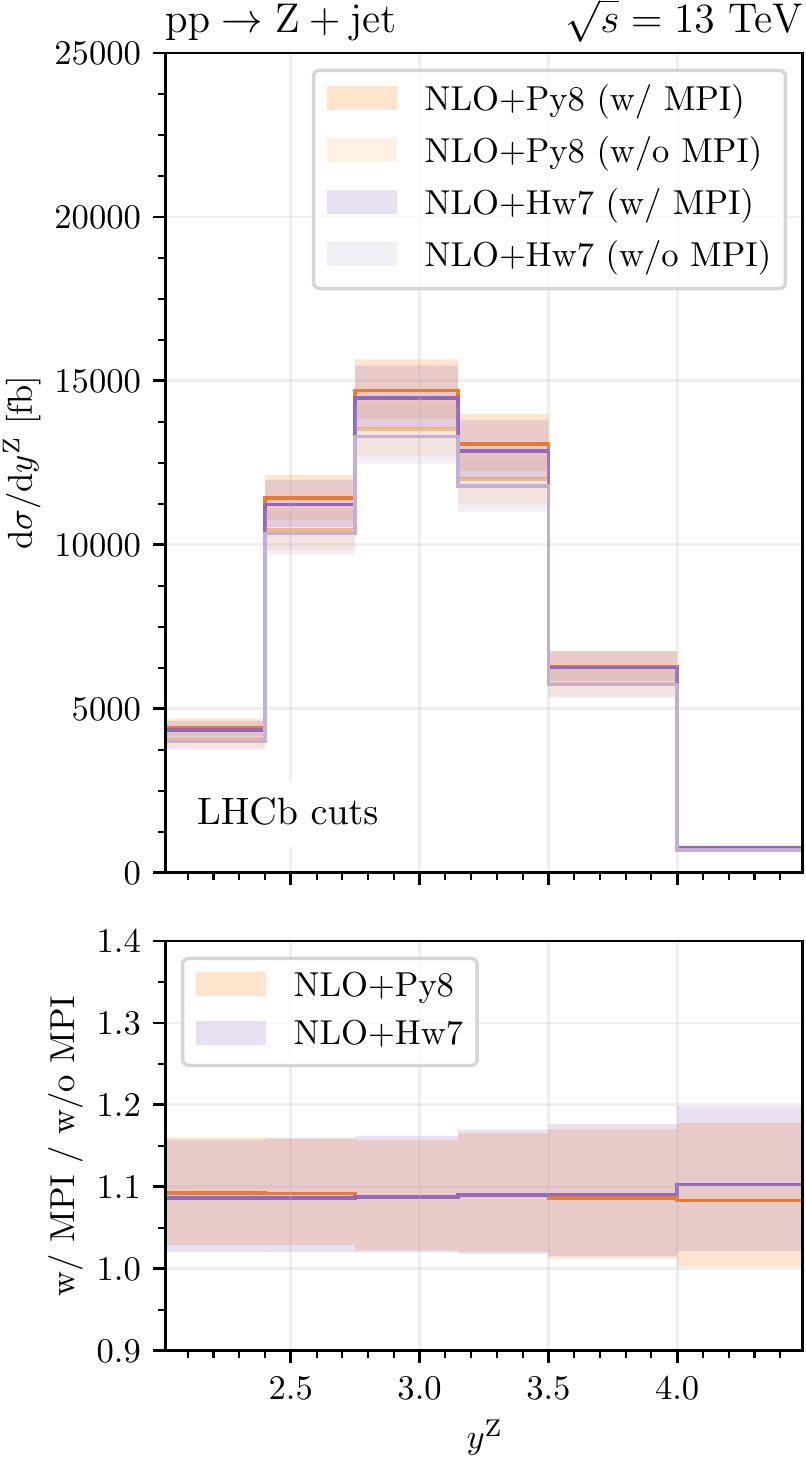}
    \subcaption{}
    \label{fig:MPI_zjet}
  \end{subfigure}
  \,\,
  \begin{subfigure}[h]{.30\textwidth}
    \includegraphics[width=\textwidth]{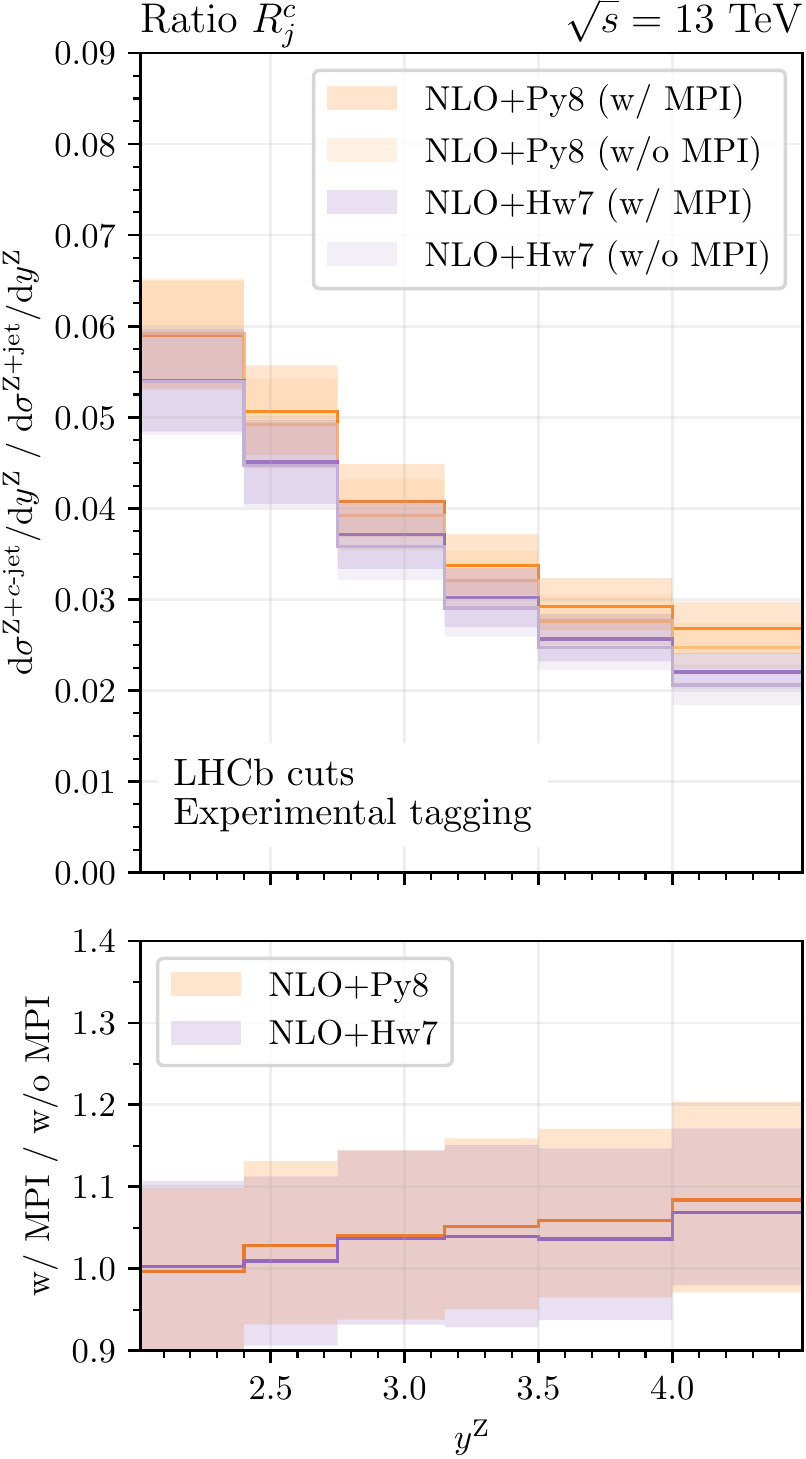}
    \subcaption{}
    \label{fig:MPI_ratio}
  \end{subfigure}  
  \caption{Effect of MPI contributions on the $\PZ$ rapidity distribution
    $y^\PZ$ in the $\PZ+\cjet$ process~(\subref{fig:MPI_zcjet}), in the
    $\PZ+\jet$ process~(\subref{fig:MPI_zjet}) and in the ratio of the
    two~(\subref{fig:MPI_ratio}). \NLOPS predictions are obtained with \PYTHIA
    (orange) or \HERWIG (purple) as parton showers. In the upper panels
    predictions including (excluding) MPI contributions are depicted in darker
    (lighter) colours. The lower panels show the ratios of curves with and
    without MPI effects.}
  \label{fig:MPI}
\end{figure*}

For the LHCb kinematics defined at the beginning of Section~\ref{subsec:set-up}, 
and also applying the (IRC unsafe) definition of jet flavour as in~\cite{LHCb:2021stx},
we observe a large contribution to the production of a $\PZ$ boson and a $\cjet$
from MPI.
In Fig.~\ref{fig:MPI_zcjet} we show the cross-section for $\PZ+\cjet$ production
after fiducial cuts, which is plotted differentially with respect to the
$\PZ$-rapidity $y^{\PZ}$. The predictions are provided at \NLOPS accuracy for
$\PZ+1j$ events generated with \mgaNLO interfaced with \PYTHIA and \HERWIG,
where the role of MPI is subsequently modelled by the two different Monte Carlo
generators.
We show the predictions obtained when including/excluding the MPI contributions,
which lead to a large (and rapidity dependent) effect on the resultant
distribution. In the central rapidity region, the effect is of order 10\%,
increasing up to 20\% at larger rapidities, and the effect is very similar
within \PYTHIA and \HERWIG.
In Fig.~\ref{fig:MPI_zjet} we further show the same plot for the unflavoured
process $\PZ+\jet$. We note a constant shift of the $y^{\PZ}$ distribution of
order 10\% due to MPI effects.
The 10\%--20\% effects observed in Fig.~\ref{fig:MPI_zcjet} for $\PZ+\cjet$
could be explained with an interplay of a constant 10\% effect (impacting both
$\PZ+\cjet$ and $\PZ+\jet$ processes) and an additional 10\% flavour-dependent
effect.
Finally, we consider the ratio $R^{c}_{j}$, even if its behaviour w.r.t. MPI
effects can be straightforwardly inferred from Fig.~\ref{fig:MPI_zcjet} and
Fig.~\ref{fig:MPI_zjet}. In particular, we observe a partial cancellation of MPI
effects, leading to a difference between predictions which is negligible at
$y^\PZ \sim 2.0$ and that increases monotonically up to 10\% at larger
rapidities. Hence, given such a partial cancellation in the ratio, the inclusion of MPI
effects seems to be particularly relevant for $3.5 \lesssim y^\PZ \lesssim
4.5$. Such a region is hugely important for the determination of the
intrinsic charm in the proton, given that the LHCb measurement in the bin
$y^\PZ \in [3.5, 4.5]$ is highly correlated to the charm PDF in the region of
the valence peak $x \simeq 0.45$~\cite{Ball:2022qks}.

As these MPI contributions arise from the overlap of multiple hard interactions, it
is not taken into account by a theoretical description based solely on single
parton scattering contributions (i.e.\ those which form the basis of a collinear
PDF fit).
If the data is to be considered for such a fit, it would be necessary to first
remove/subtract the MPI component.
Due to the complicated (overlapping) nature of multiple hard interactions, the
removal of this component may rely on theoretical modelling. As the theoretical
description of inclusive charm-quark production with the LHCb acceptance suffers
from uncertainties far in excess of $50\%$ (see for example, Fig.~6 of
Ref.~\cite{Gauld:2015yia} for $p_{\rT}^D \sim 5~\GeV$), the modelling of this
contribution (and the associated uncertainty) will require quite some care.
We note that such an uncertainty (optimistically, 50\% of the MPI contribution
as shown in Fig.~\ref{fig:MPI}) would already be in excess of the systematic
uncertainty quoted for the ratio observable in~\cite{LHCb:2021stx}.
It may be possible to overcome this issue if a data-driven based approach can be
achieved.

\end{appendices}

\bibliography{zcjet}

\providecommand{\href}[2]{#2}\begingroup\raggedright\begin{thebibliography}{10}

\bibitem{ATLAS:2018kot}
{\bf ATLAS} Collaboration, M.~Aaboud et~al., {\it {Observation of $H
  \rightarrow b\bar{b}$ decays and $VH$ production with the ATLAS detector}},
  {\em Phys. Lett. B} {\bf 786} (2018) 59--86,
  [\href{http://arxiv.org/abs/1808.08238}{{\tt arXiv:1808.08238}}].

\bibitem{CMS:2018nsn}
{\bf CMS} Collaboration, A.~M. Sirunyan et~al., {\it {Observation of Higgs
  boson decay to bottom quarks}},  {\em Phys. Rev. Lett.} {\bf 121} (2018),
  no.~12 121801, [\href{http://arxiv.org/abs/1808.08242}{{\tt
  arXiv:1808.08242}}].

\bibitem{ATLAS:2019yhn}
{\bf ATLAS} Collaboration, M.~Aaboud et~al., {\it {Measurement of VH, $
  \mathrm{H}\to \mathrm{b}\overline{\mathrm{b}} $ production as a function of
  the vector-boson transverse momentum in 13 TeV pp collisions with the ATLAS
  detector}},  {\em JHEP} {\bf 05} (2019) 141,
  [\href{http://arxiv.org/abs/1903.04618}{{\tt arXiv:1903.04618}}].

\bibitem{ATLAS:2020fcp}
{\bf ATLAS} Collaboration, G.~Aad et~al., {\it {Measurements of $WH$ and $ZH$
  production in the $H \rightarrow b\bar{b}$ decay channel in $pp$ collisions
  at 13 TeV with the ATLAS detector}},  {\em Eur. Phys. J. C} {\bf 81} (2021),
  no.~2 178, [\href{http://arxiv.org/abs/2007.02873}{{\tt arXiv:2007.02873}}].

\bibitem{ATLAS:2020jwz}
{\bf ATLAS} Collaboration, G.~Aad et~al., {\it {Measurement of the associated
  production of a Higgs boson decaying into $b$-quarks with a vector boson at
  high transverse momentum in $pp$ collisions at $\sqrt{s} = 13$ TeV with the
  ATLAS detector}},  {\em Phys. Lett. B} {\bf 816} (2021) 136204,
  [\href{http://arxiv.org/abs/2008.02508}{{\tt arXiv:2008.02508}}].

\bibitem{CMS:2019zmd}
{\bf CMS} Collaboration, A.~M. Sirunyan et~al., {\it {Search for supersymmetry
  in proton-proton collisions at 13 TeV in final states with jets and missing
  transverse momentum}},  {\em JHEP} {\bf 10} (2019) 244,
  [\href{http://arxiv.org/abs/1908.04722}{{\tt arXiv:1908.04722}}].

\bibitem{ATLAS:2021yij}
{\bf ATLAS} Collaboration, G.~Aad et~al., {\it {Search for new phenomena in
  final states with $b$-jets and missing transverse momentum in $\sqrt{s}=13$
  TeV $pp$ collisions with the ATLAS detector}},  {\em JHEP} {\bf 05} (2021)
  093, [\href{http://arxiv.org/abs/2101.12527}{{\tt arXiv:2101.12527}}].

\bibitem{ATLAS:2014jkm}
{\bf ATLAS} Collaboration, G.~Aad et~al., {\it {Measurement of the production
  of a $W$ boson in association with a charm quark in $pp$ collisions at
  $\sqrt{s} =$ 7 TeV with the ATLAS detector}},  {\em JHEP} {\bf 05} (2014)
  068, [\href{http://arxiv.org/abs/1402.6263}{{\tt arXiv:1402.6263}}].

\bibitem{LHCb:2021stx}
{\bf LHCb} Collaboration, R.~Aaij et~al., {\it {Study of Z Bosons Produced in
  Association with Charm in the Forward Region}},  {\em Phys. Rev. Lett.} {\bf
  128} (2022), no.~8 082001, [\href{http://arxiv.org/abs/2109.08084}{{\tt
  arXiv:2109.08084}}].

\bibitem{CMS:2022bjk}
{\bf CMS} Collaboration, {\it {Measurement of the production cross section of a
  W boson in association with a charm quark in proton-proton collisions at
  $\sqrt{s}=13~\mathrm{TeV}$}},  Tech. Rep. CMS-PAS-SMP-21-005, CERN, Geneva,
  2022.

\bibitem{ATLAS:2014rjv}
{\bf ATLAS} Collaboration, G.~Aad et~al., {\it {Measurement of differential
  production cross-sections for a $Z$ boson in association with $b$-jets in 7
  TeV proton-proton collisions with the ATLAS detector}},  {\em JHEP} {\bf 10}
  (2014) 141, [\href{http://arxiv.org/abs/1407.3643}{{\tt arXiv:1407.3643}}].

\bibitem{CMS:2014jqj}
{\bf CMS} Collaboration, S.~Chatrchyan et~al., {\it {Measurement of the
  production cross sections for a Z boson and one or more b jets in pp
  collisions at sqrt(s) = 7 TeV}},  {\em JHEP} {\bf 06} (2014) 120,
  [\href{http://arxiv.org/abs/1402.1521}{{\tt arXiv:1402.1521}}].

\bibitem{LHCb:2014ydc}
{\bf LHCb} Collaboration, R.~Aaij et~al., {\it {Measurement of the Z+b-jet
  cross-section in pp collisions at $ \sqrt{s} $ = 7 TeV in the forward
  region}},  {\em JHEP} {\bf 01} (2015) 064,
  [\href{http://arxiv.org/abs/1411.1264}{{\tt arXiv:1411.1264}}].

\bibitem{CMS:2017snu}
{\bf CMS} Collaboration, A.~M. Sirunyan et~al., {\it {Measurement of associated
  Z + charm production in proton-proton collisions at $\sqrt{s} = $ 8 TeV}},
  {\em Eur. Phys. J. C} {\bf 78} (2018), no.~4 287,
  [\href{http://arxiv.org/abs/1711.02143}{{\tt arXiv:1711.02143}}].

\bibitem{CMS:2016gmz}
{\bf CMS} Collaboration, V.~Khachatryan et~al., {\it {Measurements of the
  associated production of a Z boson and b jets in pp collisions at ${\sqrt{s}}
  = 8\,\text {TeV} $}},  {\em Eur. Phys. J. C} {\bf 77} (2017), no.~11 751,
  [\href{http://arxiv.org/abs/1611.06507}{{\tt arXiv:1611.06507}}].

\bibitem{ATLAS:2020juj}
{\bf ATLAS} Collaboration, G.~Aad et~al., {\it {Measurements of the production
  cross-section for a $Z$ boson in association with $b$-jets in proton-proton
  collisions at $\sqrt{s} = 13$ TeV with the ATLAS detector}},  {\em JHEP} {\bf
  07} (2020) 044, [\href{http://arxiv.org/abs/2003.11960}{{\tt
  arXiv:2003.11960}}].

\bibitem{CMS:2020hmf}
{\bf CMS} Collaboration, A.~M. Sirunyan et~al., {\it {Measurement of the
  associated production of a $Z$ boson with charm or bottom quark jets in
  proton-proton collisions at $\sqrt {s}$=13 TeV}},  {\em Phys. Rev. D} {\bf
  102} (2020), no.~3 032007, [\href{http://arxiv.org/abs/2001.06899}{{\tt
  arXiv:2001.06899}}].

\bibitem{Boettcher:2015sqn}
T.~Boettcher, P.~Ilten, and M.~Williams, {\it {Direct probe of the intrinsic
  charm content of the proton}},  {\em Phys. Rev. D} {\bf 93} (2016), no.~7
  074008, [\href{http://arxiv.org/abs/1512.06666}{{\tt arXiv:1512.06666}}].

\bibitem{Bailas:2015jlc}
G.~Bailas and V.~P. Goncalves, {\it {Phenomenological implications of the
  intrinsic charm in the $Z$ boson production at the LHC}},  {\em Eur. Phys. J.
  C} {\bf 76} (2016), no.~3 105, [\href{http://arxiv.org/abs/1512.06007}{{\tt
  arXiv:1512.06007}}].

\bibitem{Lipatov:2016feu}
A.~V. Lipatov, G.~I. Lykasov, Y.~Y. Stepanenko, and V.~A. Bednyakov, {\it
  {Probing proton intrinsic charm in photon or Z boson production accompanied
  by heavy jets at the LHC}},  {\em Phys. Rev. D} {\bf 94} (2016), no.~5
  053011, [\href{http://arxiv.org/abs/1606.04882}{{\tt arXiv:1606.04882}}].

\bibitem{Brodsky:1980pb}
S.~J. Brodsky, P.~Hoyer, C.~Peterson, and N.~Sakai, {\it {The Intrinsic Charm
  of the Proton}},  {\em Phys. Lett. B} {\bf 93} (1980) 451--455.

\bibitem{Brodsky:2015fna}
S.~J. Brodsky, A.~Kusina, F.~Lyonnet, I.~Schienbein, H.~Spiesberger, and
  R.~Vogt, {\it {A review of the intrinsic heavy quark content of the
  nucleon}},  {\em Adv. High Energy Phys.} {\bf 2015} (2015) 231547,
  [\href{http://arxiv.org/abs/1504.06287}{{\tt arXiv:1504.06287}}].

\bibitem{Ball:2022qks}
{\bf NNPDF} Collaboration, R.~D. Ball, A.~Candido, J.~Cruz-Martinez, S.~Forte,
  T.~Giani, F.~Hekhorn, K.~Kudashkin, G.~Magni, and J.~Rojo, {\it {Evidence for
  intrinsic charm quarks in the proton}},  {\em Nature} {\bf 608} (2022),
  no.~7923 483--487, [\href{http://arxiv.org/abs/2208.08372}{{\tt
  arXiv:2208.08372}}].

\bibitem{Hou:2017khm}
T.-J. Hou, S.~Dulat, J.~Gao, M.~Guzzi, J.~Huston, P.~Nadolsky, C.~Schmidt,
  J.~Winter, K.~Xie, and C.~P. Yuan, {\it {CT14 Intrinsic Charm Parton
  Distribution Functions from CTEQ-TEA Global Analysis}},  {\em JHEP} {\bf 02}
  (2018) 059, [\href{http://arxiv.org/abs/1707.00657}{{\tt arXiv:1707.00657}}].

\bibitem{Guzzi:2022rca}
M.~Guzzi, T.~J. Hobbs, K.~Xie, J.~Huston, P.~Nadolsky, and C.~P. Yuan, {\it
  {The persistent nonperturbative charm enigma}},
  \href{http://arxiv.org/abs/2211.01387}{{\tt arXiv:2211.01387}}.

\bibitem{Gauld:2020deh}
R.~Gauld, A.~Gehrmann-De~Ridder, E.~W.~N. Glover, A.~Huss, and I.~Majer, {\it
  {Predictions for $Z$ -Boson Production in Association with a $b$-Jet at
  $\mathcal {O}(\alpha_s^3)$}},  {\em Phys. Rev. Lett.} {\bf 125} (2020),
  no.~22 222002, [\href{http://arxiv.org/abs/2005.03016}{{\tt
  arXiv:2005.03016}}].

\bibitem{Czakon:2020coa}
M.~Czakon, A.~Mitov, M.~Pellen, and R.~Poncelet, {\it {NNLO QCD predictions for
  W+c-jet production at the LHC}},  {\em JHEP} {\bf 06} (2021) 100,
  [\href{http://arxiv.org/abs/2011.01011}{{\tt arXiv:2011.01011}}].

\bibitem{Hartanto:2022qhh}
H.~B. Hartanto, R.~Poncelet, A.~Popescu, and S.~Zoia, {\it
  {Next-to-next-to-leading order QCD corrections to Wbb\textasciimacron{}
  production at the LHC}},  {\em Phys. Rev. D} {\bf 106} (2022), no.~7 074016,
  [\href{http://arxiv.org/abs/2205.01687}{{\tt arXiv:2205.01687}}].

\bibitem{Gauld:2022lem}
R.~Gauld, A.~Huss, and G.~Stagnitto, {\it {A dress of flavour to suit any
  jet}},  \href{http://arxiv.org/abs/2208.11138}{{\tt arXiv:2208.11138}}.

\bibitem{Gauld:2019yng}
R.~Gauld, A.~Gehrmann-De~Ridder, E.~W.~N. Glover, A.~Huss, and I.~Majer, {\it
  {Associated production of a Higgs boson decaying into bottom quarks and a
  weak vector boson decaying leptonically at NNLO in QCD}},  {\em JHEP} {\bf
  10} (2019) 002, [\href{http://arxiv.org/abs/1907.05836}{{\tt
  arXiv:1907.05836}}].

\bibitem{Gehrmann-DeRidder:2015wbt}
A.~Gehrmann-De~Ridder, T.~Gehrmann, E.~W.~N. Glover, A.~Huss, and T.~A. Morgan,
  {\it {Precise QCD predictions for the production of a Z boson in association
  with a hadronic jet}},  {\em Phys. Rev. Lett.} {\bf 117} (2016), no.~2
  022001, [\href{http://arxiv.org/abs/1507.02850}{{\tt arXiv:1507.02850}}].

\bibitem{Banfi:2006hf}
A.~Banfi, G.~P. Salam, and G.~Zanderighi, {\it {Infrared safe definition of jet
  flavor}},  {\em Eur. Phys. J. C} {\bf 47} (2006) 113--124,
  [\href{http://arxiv.org/abs/hep-ph/0601139}{{\tt hep-ph/0601139}}].

\bibitem{Caletti:2022glq}
S.~Caletti, A.~J. Larkoski, S.~Marzani, and D.~Reichelt, {\it {A fragmentation
  approach to jet flavor}},  {\em JHEP} {\bf 10} (2022) 158,
  [\href{http://arxiv.org/abs/2205.01117}{{\tt arXiv:2205.01117}}].

\bibitem{Caletti:2022hnc}
S.~Caletti, A.~J. Larkoski, S.~Marzani, and D.~Reichelt, {\it {Practical jet
  flavour through NNLO}},  {\em Eur. Phys. J. C} {\bf 82} (2022), no.~7 632,
  [\href{http://arxiv.org/abs/2205.01109}{{\tt arXiv:2205.01109}}].

\bibitem{Czakon:2022wam}
M.~Czakon, A.~Mitov, and R.~Poncelet, {\it {Infrared-safe flavoured anti-$k_T$
  jets}},  \href{http://arxiv.org/abs/2205.11879}{{\tt arXiv:2205.11879}}.

\bibitem{Cacciari:2008gp}
M.~Cacciari, G.~P. Salam, and G.~Soyez, {\it {The anti-$k_t$ jet clustering
  algorithm}},  {\em JHEP} {\bf 04} (2008) 063,
  [\href{http://arxiv.org/abs/0802.1189}{{\tt arXiv:0802.1189}}].

\bibitem{Alwall:2014hca}
J.~Alwall, R.~Frederix, S.~Frixione, V.~Hirschi, F.~Maltoni, O.~Mattelaer,
  H.~S. Shao, T.~Stelzer, P.~Torrielli, and M.~Zaro, {\it {The automated
  computation of tree-level and next-to-leading order differential cross
  sections, and their matching to parton shower simulations}},  {\em JHEP} {\bf
  07} (2014) 079, [\href{http://arxiv.org/abs/1405.0301}{{\tt
  arXiv:1405.0301}}].

\bibitem{Sjostrand:2014zea}
T.~Sj\"ostrand, S.~Ask, J.~R. Christiansen, R.~Corke, N.~Desai, P.~Ilten,
  S.~Mrenna, S.~Prestel, C.~O. Rasmussen, and P.~Z. Skands, {\it {An
  introduction to PYTHIA 8.2}},  {\em Comput. Phys. Commun.} {\bf 191} (2015)
  159--177, [\href{http://arxiv.org/abs/1410.3012}{{\tt arXiv:1410.3012}}].

\bibitem{Bahr:2008pv}
M.~Bahr et~al., {\it {Herwig++ Physics and Manual}},  {\em Eur. Phys. J. C}
  {\bf 58} (2008) 639--707, [\href{http://arxiv.org/abs/0803.0883}{{\tt
  arXiv:0803.0883}}].

\bibitem{Bellm:2015jjp}
J.~Bellm et~al., {\it {Herwig 7.0/Herwig++ 3.0 release note}},  {\em Eur. Phys.
  J. C} {\bf 76} (2016), no.~4 196,
  [\href{http://arxiv.org/abs/1512.01178}{{\tt arXiv:1512.01178}}].

\bibitem{Bellm:2019zci}
J.~Bellm et~al., {\it {Herwig 7.2 release note}},  {\em Eur. Phys. J. C} {\bf
  80} (2020), no.~5 452, [\href{http://arxiv.org/abs/1912.06509}{{\tt
  arXiv:1912.06509}}].

\bibitem{Larkoski:2014wba}
A.~J. Larkoski, S.~Marzani, G.~Soyez, and J.~Thaler, {\it {Soft Drop}},  {\em
  JHEP} {\bf 05} (2014) 146, [\href{http://arxiv.org/abs/1402.2657}{{\tt
  arXiv:1402.2657}}].

\bibitem{PDF4LHCWorkingGroup:2022cjn}
{\bf PDF4LHC Working Group} Collaboration, R.~D. Ball et~al., {\it {The
  PDF4LHC21 combination of global PDF fits for the LHC Run III}},  {\em J.
  Phys. G} {\bf 49} (2022), no.~8 080501,
  [\href{http://arxiv.org/abs/2203.05506}{{\tt arXiv:2203.05506}}].

\bibitem{Buckley:2014ana}
A.~Buckley, J.~Ferrando, S.~Lloyd, K.~Nordström, B.~Page, M.~Rüfenacht,
  M.~Schönherr, and G.~Watt, {\it {LHAPDF6: parton density access in the LHC
  precision era}},  {\em Eur. Phys. J.} {\bf C75} (2015) 132,
  [\href{http://arxiv.org/abs/1412.7420}{{\tt arXiv:1412.7420}}].

\bibitem{Gauld:2021zmq}
R.~Gauld, {\it {A massive variable flavour number scheme for the Drell-Yan
  process}},  {\em SciPost Phys.} {\bf 12} (2022), no.~1 024,
  [\href{http://arxiv.org/abs/2107.01226}{{\tt arXiv:2107.01226}}].

\bibitem{LHCb:2015tna}
{\bf LHCb} Collaboration, R.~Aaij et~al., {\it {Identification of beauty and
  charm quark jets at LHCb}},  {\em JINST} {\bf 10} (2015), no.~06 P06013,
  [\href{http://arxiv.org/abs/1504.07670}{{\tt arXiv:1504.07670}}].

\bibitem{ATLAS:2015thz}
{\bf ATLAS} Collaboration, G.~Aad et~al., {\it {Performance of $b$-Jet
  Identification in the ATLAS Experiment}},  {\em JINST} {\bf 11} (2016),
  no.~04 P04008, [\href{http://arxiv.org/abs/1512.01094}{{\tt
  arXiv:1512.01094}}].

\bibitem{CMS:2017wtu}
{\bf CMS} Collaboration, A.~M. Sirunyan et~al., {\it {Identification of
  heavy-flavour jets with the CMS detector in pp collisions at 13 TeV}},  {\em
  JINST} {\bf 13} (2018), no.~05 P05011,
  [\href{http://arxiv.org/abs/1712.07158}{{\tt arXiv:1712.07158}}].

\bibitem{LHCb:2021dlw}
{\bf LHCb} Collaboration, R.~Aaij et~al., {\it {Identification of charm jets at
  LHCb}},  {\em JINST} {\bf 17} (2022), no.~02 P02028,
  [\href{http://arxiv.org/abs/2112.08435}{{\tt arXiv:2112.08435}}].

\bibitem{Laenen:1992zk}
E.~Laenen, S.~Riemersma, J.~Smith, and W.~L. van Neerven, {\it {Complete O
  (alpha-s) corrections to heavy flavor structure functions in
  electroproduction}},  {\em Nucl. Phys. B} {\bf 392} (1993) 162--228.

\bibitem{Riemersma:1994hv}
S.~Riemersma, J.~Smith, and W.~L. van Neerven, {\it {Rates for inclusive deep
  inelastic electroproduction of charm quarks at HERA}},  {\em Phys. Lett. B}
  {\bf 347} (1995) 143--151, [\href{http://arxiv.org/abs/hep-ph/9411431}{{\tt
  hep-ph/9411431}}].

\bibitem{Harris:1995tu}
B.~W. Harris and J.~Smith, {\it {Heavy quark correlations in deep inelastic
  electroproduction}},  {\em Nucl. Phys. B} {\bf 452} (1995) 109--160,
  [\href{http://arxiv.org/abs/hep-ph/9503484}{{\tt hep-ph/9503484}}].

\bibitem{Buza:1996wv}
M.~Buza, Y.~Matiounine, J.~Smith, and W.~L. van Neerven, {\it {Charm
  electroproduction viewed in the variable flavor number scheme versus fixed
  order perturbation theory}},  {\em Eur. Phys. J. C} {\bf 1} (1998) 301--320,
  [\href{http://arxiv.org/abs/hep-ph/9612398}{{\tt hep-ph/9612398}}].

\bibitem{Laenen:1998kp}
E.~Laenen and S.-O. Moch, {\it {Soft gluon resummation for heavy quark
  electroproduction}},  {\em Phys. Rev. D} {\bf 59} (1999) 034027,
  [\href{http://arxiv.org/abs/hep-ph/9809550}{{\tt hep-ph/9809550}}].

\bibitem{Kawamura:2012cr}
H.~Kawamura, N.~A. Lo~Presti, S.~Moch, and A.~Vogt, {\it {On the
  next-to-next-to-leading order QCD corrections to heavy-quark production in
  deep-inelastic scattering}},  {\em Nucl. Phys. B} {\bf 864} (2012) 399--468,
  [\href{http://arxiv.org/abs/1205.5727}{{\tt arXiv:1205.5727}}].

\bibitem{Gauld:2015yia}
R.~Gauld, J.~Rojo, L.~Rottoli, and J.~Talbert, {\it {Charm production in the
  forward region: constraints on the small-x gluon and backgrounds for neutrino
  astronomy}},  {\em JHEP} {\bf 11} (2015) 009,
  [\href{http://arxiv.org/abs/1506.08025}{{\tt arXiv:1506.08025}}].

\end{thebibliography}\endgroup

\end{document}